\documentclass[10pt]{iopart} %12
\usepackage{graphicx}
\usepackage[dvipsnames]{xcolor}
\usepackage{cite}

\newcommand{\blue}[1]{\textcolor{black}{#1}}
\newcommand{\magenta}[1]{\textcolor{black}{#1}}

\begin{document}

\title[]{Description of Excitonic Absorption Using the Sommerfeld Enhancement Factor and Band-Fluctuations}

\author{K. Lizárraga$^{*1}$, P. Llontop$^*$, L. A. Enrique-Morán$^*$, M. Piñeiro$^*$, E. Perez$^*$, E. Serquen$^*$, \blue{A. Tejada$^\dagger$}, F. Ruske$^\dagger$, L. Korte$^\dagger$ and J. A. Guerra$^{*,\dagger,2}$}

\address{$^{*}$Departamento de Ciencias, Sección Física, Pontificia Universidad Católica del Perú, Av. Universitaria 1801, Lima 32, Peru}
\address{$^{\dagger}$Helmholtz-Zentrum Berlin für Materialien und Energie GmbH, Division Solar Energy, Kekuléstraße 5, 12489 Berlin, Germany}
\ead{$^1$kevin.lizarraga@pucp.edu.pe, $^2$guerra.jorgea@pucp.edu.pe}
\vspace{10pt}
\begin{indented}
\item[]June 2022
\end{indented}

\begin{abstract}
One of the challenges of excitonic materials is the  accurate determination of the exciton binding energy and bandgap. The difficulty arises from the overlap of the discrete and continuous excitonic absorption at the band edge. Many researches have modeled the shape of the absorption edge of such materials on the Elliott model and its several modifications such as non-parabolic bands, magnetic potentials and electro-hole-polaron interactions. However, exciton binding energies obtained from measured data often vary strongly depending on the chosen model. Here, we propose an alternative and rather simple approach, which has previously been successful in the determination of the optical bandgap of amorphous, direct and indirect semiconductors, based on the bands-fluctuations (BF) model. In this model, the fluctuations due to disorder, temperature or lattice vibrations give rise to the well known exponential distribution of band tail states (Urbach tails). This analysis results in an analytic equation with 5 parameters only. The binding energies and optical bandgaps of GaAs and the family of tri-halide perovskites ($\textrm{MAPbX}_{3}$), $\textrm{X=Br,I,Cl}$, over a wide range of temperatures, are obtained with this model.  The results for the bandgap, linewidth and exciton binding energy are in good agreement with previous reports. Moreover, due to the polar nature of perovskites, the obtained binding energies can be compared with the ones computed with a theoretical model for polar materials via a model proposed by Kane et al.. \blue{In this model, the exciton is surrounded by a cloud of virtual phonons interacting via the Fr$\ddot{\textrm{o}}$lich interaction. As a consequence, the upper bound for the binding energy of the exciton-polaron system is calculated. Coincidentally, these results are in good agreement with the optical constants obtained with the EBF model.}
\end{abstract}

%
% Uncomment for keywords
\vspace{2pc}

% Uncomment for Submitted to journal title message
\submitto{\JPD}
%
% Uncomment if a separate title page is required
\maketitle
 
% For two-column output uncomment the next line and choose [10pt] rather than [12pt] in the \documentclass declaration
\ioptwocol

\section{Introduction}
The optical absorption of excitonic semiconductors provides information about fundamental parameters that are essential for understanding and engineering the optoelectronic technologies, such as microelectronics, optical communication and photovoltaics. \blue{Important optical parameters such as the exciton binding energy, band tail states parametrized by the Urbach energy and optical bandgap can be extracted from this part of the absorption spectrum}. For example, Urbach tail states have an impact on \magenta{devices} based on amorphous and polycristalline semiconductors \cite{guerra2,guerra3,tejada,vasudevan,asadpour}. In the same way, excitons have a profound impact in the operation mechanism of excitonic solar cells, where free charge transport is a result of the thermal dissociation of excitons into free electrons and holes \cite{baranowski}. \magenta{In the peculiar case of the planar architecture of an excitonic solar cell, such as in the ones used for perovskites, the configuration involves two carrier selective layers of wide bandgap which sandwiches the central layer of the excitonic semiconductor \cite{ball}. The energy band difference of these layers overcome the exciton binding energy forcing it to disassociate \cite{baranowski}. For this reason, knowledge of the binding energy can be helpful to engineer the excitonic solar cell architecture like the ones based on perovskites.}  

The binding energy measurement encloses experimental techniques such as optical absorption, photoluminescence (PL) and magneto-absorption. For instance, when these techniques are applied to tri-halide perovskites, they show a large range of values\cite{galkowski} like in the cases of $\textrm{MAPbBr}_{3}$  and $\textrm{MAPbI}_{3}$, where the reported values goes from $15$ to $76$ meV and $5$ to $50$ meV, respectively \cite{yang,sestu,tanaka}. At first sight, this behavior can be attributed to the different experimental conditions. However, there is a fundamental aspect of perovskites that is needed to be understood. The polar nature of perovskites defines a broad difference in the static and infinity dielectric constants \cite{soufiani,baranowski}. This, in combination with the low LO phonon energy, results in a weak electron-phonon coupling. This explanation could satisfy the large discrepancy as stated in Soufiani et al. \cite{soufiani}. However, even with this understanding of the underlying problem, we still have no tool for extracting the precise exciton binding energy. 

Usually, the procedure for extracting the binding energy from absorption measurements is based on the Elliott formula presented in eq. (\ref{elliotc}). This expression have two terms arising from the discrete and continuum absorption. Moreover, contributions from the thermal disorder and \magenta{the uncertainty principle can be modeled by the function $\mathcal{G}(z)$, presented in eq. (\ref{elliotc}), producing non-analytic expressions.} 

\begin{eqnarray}
\alpha(\hbar\omega) \approx \frac{1}{\hbar\omega} \Bigg( \sum_{n=1}^{\infty}\frac{2R^{*}}{n^{3}}& \mathcal{G}\left(E_{x_{n}}-\hbar\omega \right)  + \nonumber \\ 
&\int_{E_{g}}^{\infty}  \frac{\mathcal{G}\left(E-\hbar \omega\right) }{1-e^{-2\pi\sqrt{ \frac{R^{*}}{E-E_{g}}}}}   dE  \Bigg)\label{elliotc}
\end{eqnarray}
\begin{equation}
    E_{x_{n}}=E_{g}-\frac{R^{*}}{n^{2}} 
\end{equation}

The Elliott model is a description of the fundamental absorption when the Coulomb interaction is considered. Here, the dissipation rate energy from the photon field is completely determined by the conversion rate of photons into excitons \cite{cardona,elliott}. As a result, the transition probability per volume is written in terms of the excitonic wave functions (Wannier waves) \cite{wannier,frenkel}. These waves correspond to the well known hydrogen atom solutions for bound, $\hbar\omega \ll E_{g}$, and unbound, $\hbar\omega \gg E_{g}$, cases. The former results in discrete exciton binding energies depend on the e-h reduced mass. Whereas, the latter corresponds to a continuous range of energies above the bandgap \cite{elliott}. For energies larger than the bandgap, $\hbar\omega \gg E_g$, the continuous part is proportional to $\sqrt{\hbar\omega - E_{g}}$ and $(\hbar\omega-E_{g})^{2}$ for direct and indirect semiconductors, respectively \cite{elliott}.  Despite it success, eq. (\ref{elliotc}) shows a non-analytic expression depending on the belly distribution $\mathcal{G}(\hbar\omega)$ \cite{soufiani}. This distribution, e.g., can take the form of a Gaussian to model the effect of thermal deviations, and, a Lorentzian to account for the uncertainty principle.

In this research we propose an alternative to Elliott's equation, based on the formulation of band-fluctuations (BF) model  \cite{guerra}. The treated band-fluctuations arise from disorder-induced local variations of the band edges, thermally-induced potential fluctuations, lattice vibrations and other deviations from the perfect periodicity of the lattice \cite{guerra2,guerra}.
%The BF model works  the  density of states with a belly shaped function that carries the information of the potential fluctuations%correspondingly. 
This process results in an analytic expression (called the EBF model). \blue{Contrary to the non-analytic versions of Elliott model when convoluted with the Gaussian, Lorentzian or the hyperbolic secant distributions}. The equation, which is described in section two, contains a belly shaped peak for the discrete states, and a continuous function that grows slower than $\sqrt{\hbar\omega -E_{g}}$ for high energies, $ \hbar\omega\gg E_{g}$.

In section three we apply our EBF model to the absorption spectra of excitonic semiconductors for several temperatures. Here, we have selected GaAs, and the tri-halide perovskites based on methylammonium, $\textrm{MAPbX}_{3}$ ($\textrm{X=Pb,I,Cl}$). GaAs was selected for its well known bandgap ($E_{g}$) and exciton binding energy ($R^{*}$) for comparison purposes. And, the tri-halide perovskite family was selected due to its applications in excitonic solar cells and light-emitting diodes (LEDs) \cite{marongiu}, as well as due to the intriguing variability of the binding energy caused by their polar nature \cite{baranowski,soufiani}. With this materials, we aim to test the applicability of the EBF model for excitonic, polar and non-polar, materials. \blue{Nonetheless, this analysis can be extended to other polar excitonic materials like the organic compounds and the Transition Conductive Oxides (TCOs) like AZO.}

The evolution of the parameters with temperature can be found in section three. In the case of the bandgap, this behavior can set the value of the LO phonon energy by using the Varshni \cite{varshni}, Passler \cite{passler}, Viña \cite{vina} or the two oscillator models based on the electron-phonon contribution \cite{gobel,mannino,street}. However, there are several reports in which the thermal expansion term is no negligible when describing the bandgap evolution, like in the case of perovskites \cite{wei,francisco2019,yu2011}. \blue{The complete model considering both contributions was developed in several studies of Cardona et al. and co-authors \cite{cardona2,gobel,Lautenschlager,serrano}. In this work, we develop an analysis regarding the consideration of both contributions and for each contribution independently} for GaAs and tri-halide perovskites. This analysis concludes with the extraction of the LO phonon energy of GaAs and the tri-halide perovskites.

Another method for extracting the LO phonon energy is through the broadening of the excitonic peak. This is described in terms of the Full Width Half Maximum (FWHM), also called linewidth. The thermal evolution of this quantity, described by the works of Toyozawa et al. \cite{toyozawa2,segall}, is composed by the coupling between the acoustic and optical phonons with electrons. For our analyzed semiconductors, the latter coupling is the dominant. As a consequence, the linewidth is proportional to the weak electron-phonon coupling, a.k.a, Fr$\ddot{\textrm{o}}$lich interaction. This causes a line shape proportional to the boson occupation number which is used to determine the LO phonon energies for tri-halide perovskites. In the peculiar case of the tri-halide perovskites, the obtained LO phonon energies of the bandgap and linewidth analysis are important for analyzing the large variability of exciton binding energy values. 

The theoretical description of the absorption of excitons in a polar semiconductor is described in the works of Pollmann and Kane \cite{pollmann,kane}. \blue{In their model, the system is described as the interaction of excitons and phonons through the Fr$\ddot{\textrm{o}}$lich interaction. The upper limit for the binding energy can be settle by applying a set of canonical transformations to the Hamiltonian. In this way, the contributions of the electron-phonon interaction are separated, allowing the use of the exciton wave function as in the hydrogen-like model. As a result, the upper bound for the binding energy is written in a set of self-consistent equations that depend on the LO phonon energy. The application of Kane's model for the tri-halide perovskites give an excellent agreement when compared with our binding energy results of the EBF model.}

\section{Theory and Methodology}
\subsection{Brief summary of the Elliott model}
The Elliott model describes the exciton as a hydrogen atom. In this picture, the electron is excited from the valence to the conduction band leaving a hole behind. They interact trough the Coulomb potential producing a quasi-particle known as exciton. \blue{This produces new energy levels in the region between the conduction and valence bands. As a consequence, the exciton system absorbs light for energies below the bandgap}. These states correspond to the bound solutions. On the other hand, the unbound states \magenta{are formed} when the photon energy is larger than the band edge \cite{cardona,elliott,murayama}. The optical transition rate per unit volume and per unit time for converting a photon into an exciton is given by:
\begin{equation}
R_{cv}=\frac{2\pi}{\hbar}\sum_{f} |\langle f \vert \mathcal{H}_{xR} \vert 0 \rangle |^{2} \delta(E_{f}(K)-E_{0}-E).
\end{equation}
Here, \blue{the photon energy is $E=\hbar\omega$}, the exciton-photon interaction is expressed by $\mathcal{H}_{xR}$, the initial state with no excitons is $\vert 0 \rangle$, and, the final state is $\langle f \vert$. This state represents an exciton with energy $E_{f}$ and wave vector $K$:
\begin{equation}
E_{f}(K)= \frac{\hbar K^{2}}{2\mu} - \frac{R^{*}}{n^{2}} \approx  -\frac{R^{*}}{n^{2}}, \label{eq.ef}
\end{equation}
Here $\mu$ is the electron-hole reduced mass (exciton mass), $R^{*}$ is the exciton Rydberg constant in terms of $\mu$, $R^{*}=\mu e^{4}/2\hbar^{2} (4\pi\varepsilon_{0})^{2}$, and $n$ is the exciton energy level. The wave vector $K$ is the sum of the electron $k_e$ and hole $k_h$. And, in the case of direct transitions, \blue{neglecting the photon momentum}, the wave vector conservation is $k_e=-k_h=k$. This produces a total wave vector of $K=0$ \cite{elliott} as seen in the right hand of eq. (\ref{eq.ef}). As a consequence, the final energy depends only on the n$^{th}$ discrete state. 

For optical transitions, we have the allowed and forbidden transitions depending on the nature of the hydrogen wavefunction. \blue{On one hand, the allowed transitions are the ones we observe in the optical measurements and corresponds to the s-like states. On the other hand, the forbidden transitions, described by p-like states, are usually called dark excitons due to its \magenta{minuscule amplitude in the optical spectrum. This is a consequence of the amplitude being inversely proportional to the excitonic radius which is of the size of several unit cells} \cite{wannier}. For this reason, our study is focused on the allowed transitions that shape the absorption spectrum.} The interaction potential for these transitions, $\langle f \vert \mathcal{H}_{xR} \vert 0 \rangle$, can be expressed in terms of the conduction ($\psi^{c}_{k}$) and valence ($\psi^{v}_{k}$) Bloch wave-functions, i.e.,
\begin{equation}
|\langle f \vert \mathcal{H}_{xR} \vert 0 \rangle|^{2} =NV |\phi_{nlm}(0)|^{2} |\langle \psi^{c}_{k} \vert \mathcal{H}_{eR} \vert\psi^{v}_{k}\rangle|^{2}. \label{hint}
\end{equation}

Eq. (\ref{hint}) states that the probability of exciting an exciton optically is proportional to the overlap of the electron and hole wavefunctions. Here $N$ is the number of unit cells. $V$ is the volume of a unit cell. $\phi_{nlm}(r)$ is the analytic solution for two spherical single bands with effective masses $m_e$ and $m_h$ \cite{elliott}. This imply that the hydrogen atom solutions can describe the new final states, $f$. As a consequence, the solution is in terms of the coefficients $n,l,m$ which are the principal, orbital and magnetic quantum numbers. And, the opposite momentum of the electron and hole produces solutions independent of the relative position, $\phi_{nlm}(r=0)$. Lastly, the electric interaction $\mathcal{H}_{eR}$ follows the behavior of a dipole \cite{elliott}, i.e.,
\begin{equation}
|\langle \psi^{c}_{k} \vert \mathcal{H}_{eR} \vert \psi^{v}_{k} \rangle|^{2} = \left( \frac{e|\vec{E}|}{2m_{e}\omega} \right)^{2}|\langle \psi^{c}_{k} \vert \hat{\textbf{e}}.\vec{\textbf{p}} \vert \psi^{v}_{k} \rangle|^{2}, \label{xr_to_er}
\end{equation}
with $e$ and $m_e$ being the electron charge and effective mass, $\omega$ is the \blue{photon} frequency, $\vec{E}$ is the external electric field amplitude, and $\hat{e}$ is the unit vector of the polarization. 

For the term $\phi_{nlm}(0)$ in eq. \ref{hint}, the only non-vanishing expressions are the $S$-like states ($l,m=0$). Thus, implying that excitons will have $S$ symmetry when formed. For energies below the bandgap, the bound allowed solutions for the hydrogen atom are:
\begin{equation}
    |\phi_{n00}(0)|^2=1/(NV\pi(a_{0}n)^{3}).
\end{equation}
Here $a_{0}$ is the Bohr radius for an exciton with reduced mass $\mu^{-1}=m_e^{-1}+m_h^{-1}$.

The transition rate for exciton formation is
\begin{equation}
R^{d}_{cv} = \mathcal{R}_{d} |P_{cv}|^{2} \sum_{n=1}^{\infty} \frac{1}{n^{3}}\delta(E  - E_{x_{n}}), \label{Elliottd} 
\end{equation}
with
\begin{equation}
\mathcal{R}_{d} = \frac{\mu^{3}e^{8}|\vec{E}|^{2}}{m_{e}^{2}\omega^{2}\hbar^{7}(4\pi\epsilon_{0})^{3}}.
\end{equation} 
\blue{The exciton peak location is $E_{x_{n}}=E_{g}-\frac{R^{*}}{n^{2}}$, and the matrix element is $P_{cv}=\langle \phi^{c}_{k} \vert \hat{\textbf{e}}.\vec{\textbf{p}} \vert \phi^{v}_{k} \rangle$}. In addition, a factor of two has been included to count the spin degeneracy. The super/sub-script $d$ has been added to refer to the discrete scenario.

For the continuous scenario, the employed solutions will correspond to the unbound energy states of the hydrogen atom written in terms of the hyper-geometric wavefunctions, $R(r)$ \cite{cardona,shinada}. As in the previous case, we pick the $r=0$ solution, i.e.,
\begin{equation}
R(0)= \kappa\pi \frac{ e^{\pi\kappa}}{(NV)^2 \textrm{sinh}(\pi\kappa)}, 
\end{equation}
where \blue{the constant $\kappa$ is} $\sqrt{R^{*}/(E-E_{g})}$. The interaction term can be written as:
\begin{equation}
|\langle f \vert \mathcal{H}_{xR} \vert 0 \rangle|^{2}= \left( \frac{e|\vec{E}|}{2m_{e}\omega} \right)^{2} \frac{\kappa\pi e^{\pi\kappa}}{NV \textrm{sinh}(\pi\alpha)} |P_{cv}|^{2}.
\end{equation}

On the other hand, the energy difference of the final and initial states, $E_{f}(K)-E_{0}$, becomes the difference from the valence and conduction band, $E_{c}-E_{v}=E_{cv}$. As a consequence, the sum over the continuum of energies transform the discrete summation into an integral, in terms of the Density Of States (DOS) for direct transition semiconductors, $D_{cv}$, i.e., 
\begin{equation}
    \sum_f \delta(E_{cv}-E) \rightarrow \int NV  D_{cv}(E_{cv})\delta(E_{cv}-E)dE_{cv},
\end{equation}
with
\begin{equation}
D_{cv}(E_{cv})=\frac{\sqrt{2}\mu^{3/2}}{\pi^{2}\hbar^{3}}\sqrt{E_{cv}-E_{g}}\Theta(E_{cv}-E_{g}). \label{directjdos}
\end{equation}

Thus, the transition rate for the continuous case (c) is:
\begin{equation}
R^{c}_{cv}=\frac{2\pi}{\hbar} \left( \frac{e|\vec{E}|}{2m_{e}\omega} \right)^{2} \frac{ \kappa \pi e^{\pi\kappa}}{\textrm{sinh}(\pi\kappa)} |P_{cv}|^2 D_{cv}(E),
\end{equation}
and by substituting eq. (\ref{directjdos}),
\begin{eqnarray}
R^{c}_{cv}&=\mathcal{R}_{c} |P_{cv}|^{2} \frac{ \kappa \sqrt{E -E_{g}} e^{\pi\kappa}}{\textrm{sinh}(\pi\kappa)}, \\
&= 2\mathcal{R}_{c} |P_{cv}|^{2}R^{*}\frac{1}{1-e^{-2\pi\kappa}}, \label{elliotcc2} 
\end{eqnarray}
with,
\begin{equation}
\mathcal{R}_{c} =\frac{\sqrt{2}\mu^{3/2}}{2\hbar^{4}} \left( \frac{e|\vec{E}|}{m_{e}\omega} \right)^{2}. 
\end{equation}
From eqs. (\ref{Elliottd}) and (\ref{elliotcc2}), \blue{we can write the expression for the absorption given the relation between the transition rate and imaginary part of the dielectric constant:
\begin{equation}
\varepsilon_{i} =\frac{8\pi\hbar R_{cv}}{|\vec{E}|^{2}}. \ \ 
\end{equation}
Thus, the absorption coefficient is:
\begin{equation}
\alpha=\frac{\omega\varepsilon_{i}}{4\pi\varepsilon_{0}\hat{n} c}=\frac{2\hbar\omega R_{cv}}{\varepsilon_0 |\vec{E}|^2 \hat{n} c}.
\end{equation}}
Here, $\hat{n}$ is \blue{the real part} of the refractive index, and $c$ is the speed of light. The total absorption coefficient is the sum of the discrete and continuum parts. This is left in terms of the exciton Rydberg constant for fitting purposes. 
\begin{equation}
\alpha(E)=\alpha_{d}(E)+\alpha_{c}(E), \label{simple.sum}
\end{equation}
\begin{eqnarray}
\alpha(E)=\frac{\mathcal{A}}{E} \Bigg( 2R^{*}
\sum_{n} \frac{1}{n^{3}}\delta(E& -E_{x_{n}}) + \nonumber \\
&\frac{1}{1-e^{-2\pi\sqrt{\frac{R^*}{E-E_g}}}} \Bigg). \label{alphadc}
\end{eqnarray}
with
\begin{eqnarray}
\mathcal{A}&=\frac{8\hbar(4\pi\varepsilon_{0})^{3}R^{*^{2}}}{\varepsilon_0  e^{4}m_{e}^{2} \hat{n}c}|P_{cv}|^{2}. \label{constA}
\end{eqnarray}

Eq. (\ref{alphadc}) is compatible with other reports \cite{elliott,sell}. In addition, it is commonly convoluted with a  belly shaped function $\mathcal{G}(E-\hbar\omega)$ for adding the thermal and uncertainty principle contributions. For instance, a Gaussian and Lorentizan are used in Soufiani et al. \cite{soufiani}, whilst a hyperbolic secant is used in Saba et al. \cite{saba}. The final version of the Elliott equation  can be expressed as \cite{sestu,yang,fabian}:
\begin{eqnarray}
\alpha(\hbar\omega)=\frac{\mathcal{A}}{\hbar\omega} \Bigg( 2R^{*}\sum_{n} &\frac{1}{n^{3}} \mathcal{G}\left( E_{x_{n}} - \hbar\omega \right) + \nonumber \\
&\int_{E_{g}}^{\infty} \frac{\mathcal{G}\left(E- \hbar\omega \right)}{1-e^{-2\pi\sqrt{ \frac{R^{*}}{E-E_{g}}} }}dE \Bigg). \label{alphadcg}
\end{eqnarray}

\blue{This convolution process is similar to the band-fluctuations model performed for the optical absorption of non-excitonic materials. In the BF model, a convolution can be applied to the density of states for producing the tail states of the Urbach region. Likewise, the BF model can be applied  directly to the absorption coefficient producing a similar expression to eq. (\ref{alphadcg}) for non-excitonic materials. In the BF model, the fluctuations around the bandgap are attributed to disorder, thermal effects or lattice deformations. These are modeled in terms of a Gaussian \cite{john,oleary2,oleary3} or by using a distribution inspired on the derivative of the Fermi distribution \cite{guerra}. In this work, we follow the latter due to its success for explaining the absorption edge of direct, indirect and amorphous non-excitonic semiconductors \cite{guerra,guerra2,guerra3}}

\subsection{Elliott-BF model}
\blue{The idea of potential fluctuations were proposed initially to model a semi-classical density of states for heavily doped semiconductors \cite{kane}. This formalism were applied to the optical absorption, by using Gaussian distribution, for the description of the exponential Urbach tails \cite{john}. Later, O'leary et al. deduce an expression for the optical absorption in amorphous materials \cite{oleary2,oleary3}. And, Guerra et al. use an alternative distribution for modeling the absorption edge on direct, indirect and amorphous semiconductors. This was recently extended for modeling high absorption energy absorption regions in the work of Lizarraga et al \cite{lizarraga}. In the present work, we model the fluctuations accordingly with the distribution given in Guerra et al., i.e., the distribution $\mathcal{G}(z)$ will acquire the form:}
\begin{equation}
    \mathcal{G}_{BF}(z)=\frac{1}{\sigma}\frac{ e^{z/\sigma}}{(1+e^{z/\sigma})^2}. \label{w}
\end{equation}
Here, $\sigma$ is the belly width. \blue{In the predecessor case, the obtained expressions for the absorption coefficient of non-excitonic direct and indirect materials are non-analytic. These are written in terms of the $\textrm{polylogarithmic}$ functions, which are products of the Fermi integrals that are computable numerically.} They produces excellent results when contrasted with experiments as shown in the works of Guerra and co-workers \cite{guerra,guerra2,guerra3}. Following this success, we applied eq. (\ref{w}) to the transition rate of excitonic materials. 

Since the absorption in the presence of the Coulomb interaction has two contributions, it is important to remark that in the case of the discrete absorption, the broadening affects the exciton energy levels. Correspondingly, in the continuous case, \blue{these fluctuations act on the mobility edge further affecting the $E_{g}$ determination.} The discrete transition rate, eq. (\ref{Elliottd}), is now:
%We state a possibly origin of the fluctuations regarding the the excitons (who is gonna be tested later in this research): due to the lattice vibrations, polarons can be formed when the phonons interact strongly with electrons as in the case of perovskites whose has a large Fr$\ddot{o}$lich coupling constant \cite{baranowski}, thus causing the fluctuations to carry this polaronic contribution.
\begin{eqnarray}
\langle R^{d}_{cv} \rangle= \mathcal{R}_{d}|P_{cv}|^{2} \sum_{n=1}^{\infty} \int_{E_{x_{n}}}^{+\infty} \frac{1}{n^{3}} &\delta(E-E_{x_{n}}) \times \nonumber \\
&\mathcal{G}_{BF}(E-\hbar\omega)dE.
\end{eqnarray}
Yielding to the following result:
\begin{equation}
\langle R^{d}_{cv} \rangle=\mathcal{R}_{d}|P_{cv}|^{2} \sum_{n=1}^{\infty} \frac{1}{n^{3}} \mathcal{G}_{BF}(E_{x_{n}}-\hbar\omega). \label{eq.trans.disc}
\end{equation}
For the transition rate of the continuum, eq. (\ref{elliotcc2}), we have:
\begin{eqnarray}
\langle R^{c}_{cv} \rangle= \mathcal{R}_{c}|P_{cv}|^{2} \int_{E_{g}}^{+\infty} \frac{\mathcal{G}_{BF}(E-\hbar\omega)}{1-e^{-2\pi\sqrt{ \frac{R^{*}}{E-E_{g}}} }} dE, \label{convolcint}
\end{eqnarray}
\blue{which, after following the steps presented in the Appendix, produces an analytic expression:}
\begin{eqnarray}
\langle R^{c}_{cv} \rangle=2\sqrt{R^{*}}\mathcal{R}_{c}|P_{cv}|^{2} \Bigg( &\frac{1}{1+e^{(E_{g}-\hbar\omega)/\sigma}}+ \nonumber\\
&\frac{1}{e^{2\pi\sqrt{\frac{R^{*}}{\hbar\omega-E_{g}}}}-1} \Bigg). \label{eq.trans.cont}
\end{eqnarray}

\blue{With eqs. (\ref{eq.trans.disc}) and (\ref{eq.trans.cont}), we can write the  expression for the absorption coefficient following eq. (\ref{simple.sum}). This will be denoted as $\alpha_{EBF}$ and have the form:}
\begin{eqnarray}
\alpha_{EBF}(\hbar\omega)=&\frac{\mathcal{A}}{\hbar\omega} \Bigg( 2R^{*}\sum_{n} \frac{1}{n^{3}} \mathcal{G}_{BF}(E_{x_{n}}-\hbar\omega) + \nonumber \\ 
&\frac{1}{1+e^{(E_{g}-\hbar\omega)/\sigma}}+\frac{1}{e^{2\pi\sqrt{\frac{R^{*}}{\hbar\omega-E_{g}}}}-1} \Bigg). \label{alphadc2}
\end{eqnarray}
Here $\mathcal{A}$ is the constant previously found on eq. (\ref{constA}). The main features of eq. (\ref{alphadc2}) are its analytic form and the dependency on just 4 parameters. These are, the constant ($\mathcal{A}$), the bandgap ($E_g$), the belly width ($\sigma$), and the exciton binding energy ($R^{*}$). 

\subsection{\magenta{Pseudo-Voigt Profile}}
The computed absorption of the EBF model arises from considering only thermally induced fluctuations.  However, when  experimental techniques are applied, the uncertainty principle produces a contribution that is modeled as a Lorentzian in optic related experiments. In this section we review a pseudo-Voigt procedure for the inclusion of two different broadenings, which will be helpful for the extension of our model to embrace the uncertainty principle contribution.

In 2015, Soufiani et al. propose a pseudo-Voigt profile for the Lorentzian and Gaussian profiles \cite{soufiani}. In their work, the broadenings were not attributed directly to a physical meaning, despite establishing that the Lorentzian profile contribution increases when temperature raises. They arrive at an analytical expressions, after the convolution process, achieved by approximations \cite{soufiani2}. Thus, the absorption coefficient looks like:
\begin{equation}
    \alpha(\hbar\omega)=\eta \alpha_{G}(\hbar\omega) + (1-\eta)\alpha_{L}(\hbar\omega), \label{alphasouf}
\end{equation}
where the Gaussian broadening is described by,
\begin{eqnarray}
    \alpha_{G}=&\frac{A\sqrt{R^*}}{\hbar\omega} \Bigg( 2R^{*} \sum_{n=1}^{N} \frac{e^{-\left( \hbar\omega-E_{g}+\frac{R^{*}}{n^{2}} \right)^{2}/2\tau_{n}^2}}{n^{3}\sqrt{2\pi}\tau_{n}} + \nonumber\\
    &\frac{1+\textrm{erf}\left(\frac{\hbar\omega-E_{g}}{\sqrt{2}\tau_{c}} \right)}{2} + \frac{\tau_{c}e^{-\left(\hbar\omega-E_{g}-R^{*} \right) ^{2}/2\tau_{c}^2}}{58\sqrt{2\pi}} +\nonumber \\
    &\frac{\hbar\omega-E_{g}-R^{*}}{116 R^{*}} \times \nonumber \\
    &\left( 1+\textrm{erf} \left( \frac{\hbar\omega-E_{g}-R^{*}}{\sqrt{2}\tau_{c}} \right) \right)  \Bigg), \label{alphagsouf}
\end{eqnarray}
and the Lorentzian broadening is,
\begin{eqnarray}
    \alpha_{L}=&\frac{A\sqrt{R^*}}{\hbar\omega} \Bigg( \frac{2R^{*}}{\pi} \sum_{n=1}^{N} \frac{\Gamma_{n}/n^3}{(\hbar\omega-E_{g}+\frac{R^{*}}{n^{2}} )^2 + \Gamma_{n}^2} + \nonumber \\
    &\left[ 0.5 + \textrm{arctan} \left(  \frac{\hbar\omega-E_{g}}{\Gamma_{c}}\right)/\pi \right] + \nonumber \\
    & \textrm{arctan} \left( \frac{\hbar\omega-E_{g}-R^*}{\Gamma_{c}} \right)\frac{(\hbar\omega-E_{g}-R^*)}{58\pi R^*} - \nonumber \\
    & \textrm{arctan}\left( \frac{\hbar\omega-E_{m}}{\Gamma_{c}} \right) \left( \frac{\hbar\omega-E_{m}}{58\pi R^*} \right) + \nonumber \\ &\frac{E_{m}-E_{g}-R^*}{116 R^*} - \frac{\Gamma_{c}}{116\pi R^* } \times \nonumber \\
    &\textrm{ln} \Bigg[ \frac{(E-E_{g}-R^*)^2 + \Gamma_{c}^2}{(\hbar\omega-E_{m})^2 + \Gamma_{c}^2} \Bigg]. \label{alphalsouf}
\end{eqnarray}

Here, $\eta$ in eq. (\ref{alphasouf}) measures the weight of each contribution. $\tau$ and $\Gamma$ in eqs. (\ref{alphagsouf}) and (\ref{alphalsouf}), respectively, have units of energy and represent the broadening of the Gaussian and Lorentzian. The subscript $n$ and $c$ has been added to denote the discrete and continuum, respectively, for each profile. For instance, in the case of eq. (\ref{alphagsouf}), $\tau_{1}$ defines the broadening of the first discrete peak, while $\tau_{c}$ does it for the continuum. \blue{Higher order broadening discrete peaks} follow the Toyozawa relation \cite{soufiani,toyozawa}:
\begin{equation}
    \tau_{n}=\tau_{c}-\frac{\tau_{c}-\tau_{1}}{n^2}.\label{eq.sigman}
\end{equation}
Likewise, the broadening $\Gamma$ follow the same relations of eq. (\ref{eq.sigman}). In addition, the two broadenings, $\Gamma$ and $\tau$, are correlated \blue{(see eq. (\ref{eq.correlated}))} by sharing the same Full Width Half Maximum (FWHM) according to the procedure of Soufiani et al. \cite{soufiani2}. \blue{This allows an improvement of the fitting region and  the reduction of one free parameter. 
\begin{equation}
    2\Gamma_{n}=2\sqrt{2\textrm{ln}(2)}\tau_{n}. \label{eq.correlated}
\end{equation}}

\subsection{\magenta{Uncertainty Principle in the EBF model}}
\blue{The pseudo-Voigt profile, performed by Soufiani et al., described in the previous section is applied to improve our EBF model.} Here, we describe the absorption to be composed of an EBF profile, accounting for the thermal contribution, and a Lorentzian profile, for the uncertainty principle contribution. This reads as:
\begin{equation}
    \alpha(\hbar\omega)=\eta \alpha_{EBF}(\hbar\omega) + (1-\eta)\alpha_{L}(\hbar\omega), \label{alphaeta}
\end{equation}
with
\begin{eqnarray}
    \alpha_{EBF}&= \frac{A_{1}\sqrt{R^*}}{\hbar\omega} \times \nonumber \\
    &\Bigg( 2R^{*}\sum_{n=1}^{N} \frac{1}{n^{3}\sigma_{n}} 
    \frac{e^{- \left( \hbar\omega-E_{g}+\frac{R^*}{n^2} \right)/\sigma_{n}}}{\left( 1+e^{- \left( \hbar\omega-E_{g}+\frac{R^{*}}{n^2} \right) /\sigma_{n}} \right)^2} + \nonumber \\ &A_{2} \left( \frac{1}{1+e^{(E_{g}-\hbar\omega)/\sigma_{c}}}+ \frac{1}{e^{2\pi\sqrt{\frac{R^{*}}{\hbar\omega-E_{g}}}}-1} \right) \Bigg), \label{alphadc3}
\end{eqnarray}
whereas $\alpha_{L}$ is described by eq. (\ref{alphalsouf}). 

Equation (\ref{alphadc3}) is written in terms of the different broadenings for the discrete and continuous parts. \blue{Note that the discrete peaks follow the relation of eq. (\ref{eq.sigman}). And, the correlation between the Lorentzian and EBF FWHM is $2\Gamma_{n}=4\sigma_{n}\textrm{arccosh}(\sqrt{2})$.} 

It is remarkable to say that during the implementation of eq. (\ref{alphadc3}), the different coefficients, $A_{1}$ and $A_{2}$, has been used for the discrete and continuous, respectively. This difference can be understood from the matrix element which can be modeled as Lorentz oscillator for regions above the bandgap \cite{modine,ferlauto}. \blue{In addition, differences between eqs. (\ref{alphadc3}) and (\ref{alphadc2}), such as the extra $\sqrt{R^*}$ alongside $A_{1}$, are related to our inclination of constructing an expression similar to the ones used by Soufiani et al. \cite{soufiani2}.} We highlight that eq. (\ref{alphadc3}) have six parameters only. Those are, $A1$, $A2$, $\sigma_{d}$, $\sigma_{c}$, $E_{g}$ and $E_{b}$. 

Lastly, it is remarkable that during the fitting procedure with eq. (\ref{alphaeta}), the best fitted parameters were achieved in the regime of $\eta\approx0.99$. Thus, indicating that only the thermal effects, modeled with the EBF model of eq. (\ref{alphadc3}), contribute to shape the absorption for the materials analyzed in this work. Nonetheless, this behavior may not be generalized for other materials, and the full expression of eq. (\ref{alphaeta}) must be used.

\section{Results}
We now apply eq. (\ref{alphaeta}) \blue{with $\eta\approx0.99$, or equivalently, eq. (\ref{alphadc3}),} to describe the excitonic absorption of GaAs, and tri-halide perovskites, $\textrm{MAPbBr}_{3}$, $\textrm{MAPbI}_{3}$ and $\textrm{MAPbI}_{3-x}\textrm{Cl}_{x}$. \blue{The spectral data of GaAs versus T was extracted from Sturge \cite{sturge} with sample temperatures of} 21K, 90K, 186K and 294K. The optical spectrum data of $\textrm{MAPbBr}_{3}$ and $\textrm{MAPbI}_{3}$, over the ranges of temperatures from 7K to 300K, were extracted from Soufiani et al. The data of $\textrm{MAPbI}_{3-x}\textrm{Cl}_{x}$ were extracted from D'Innocenzo et al. \cite{dinn}. The sets of data of the perovskites were adjusted linearly in order to remove the bias of the transmittance measurements from the original works. \magenta{It is important to remark that the concentration of the $\textrm{MAPbI}_{3-x}\textrm{Cl}_{x}$ is not determined in the original work of D'Innocenzo et al. \cite{dinn}. Nonetheless, we can infer its concentration by looking into the optical spectrum (see figure \ref{fig:abs_coeff_mapi_mapcl}c) which has similar energy ranges as the $\textrm{MAPbI}_{3}$. Thus, implying that the final concentration should be more favorable to $\textrm{MAPbI}_{3}$ than to $\textrm{MAPbCl}_{3}$}

In this section, our results are divided in the following. We start by describing the absorption curves and the fitting procedure. Then, we analyze the evolution of parameters such as bandgap, FWHM and binding energy, with sample temperature. \magenta{The evolution of the bandgap is analyzed with the model of Cardona et al. \cite{cardona} and Varshni et al. \cite{varshni}, while the FWHM is described in terms of the model of Toyozawa et al. \cite{toyozawa2}. Lastly, the evolution of the binding energy is described theoretically with the model of Pollmann and Kane \cite{pollmann,kane}}.

\subsection{Absorption \blue{Fitting}}

\begin{figure*}[!htb]\centering % Using \begin{figure*} makes the figure take up the entire width of the page
\includegraphics[scale=0.29]{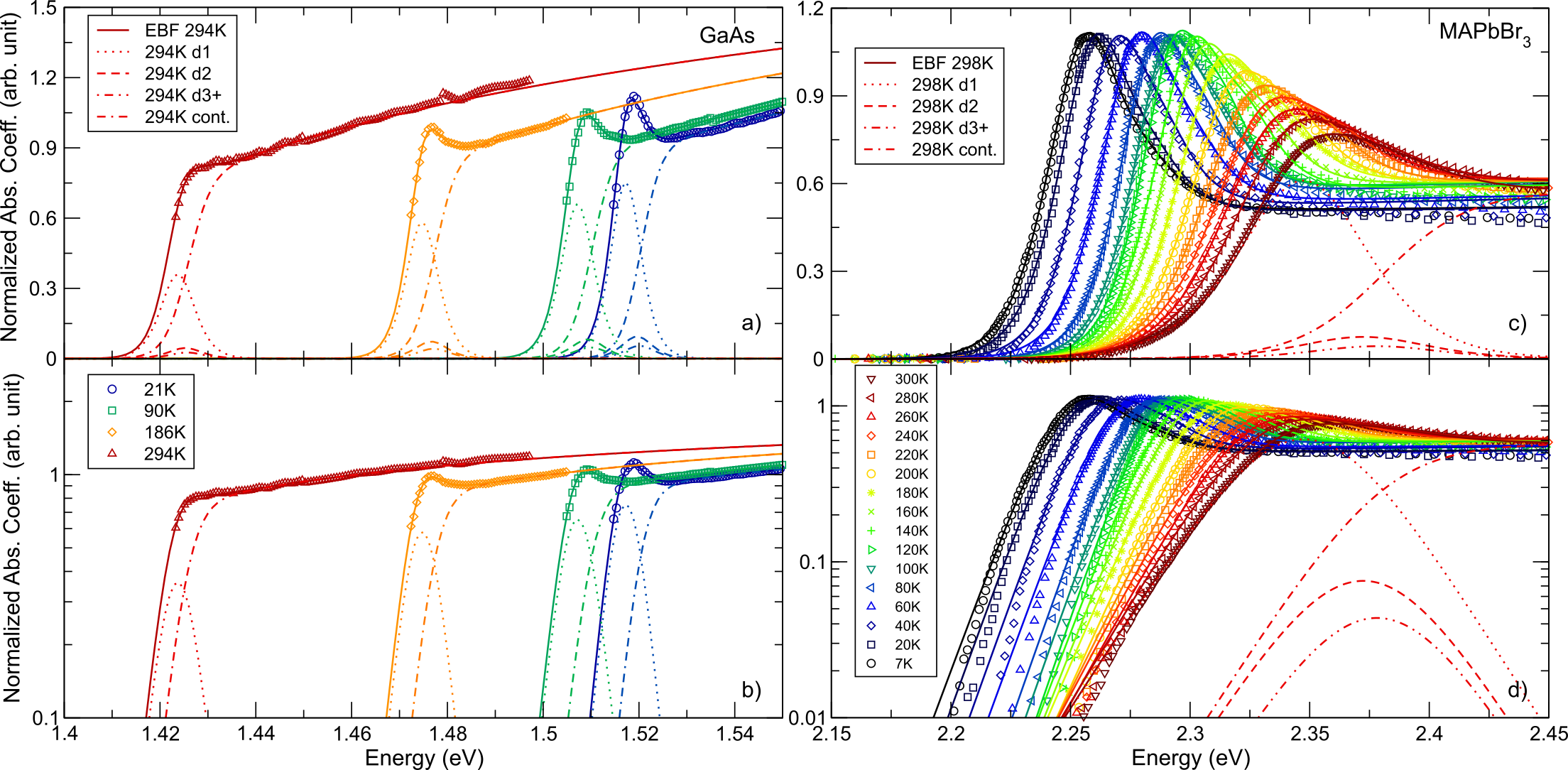}
\caption{Normalized absorption coefficient of GaAs, extracted from Sturge \cite{sturge}, fitted with the EBF model for several temperatures in linear (a) and logarithmic (b) scale. Here the contributions of the discrete peaks and the continuum are plotted in dashed lines in figures (a) and (b). Normalized absorption coefficient of $\textrm{MAPbBr}_{3}$, extracted from Soufiani et al. \cite{soufiani}, fitted with the EBF model for several temperatures in linear (c) and logarithmic scale (d). The contributions of the discrete and continuum parts of the spectra at T=298K are shown in red dashed lines, only, for viewing reasons. \blue{Here the excitonic peaks are denoted as ``d1'', ``d2'' and ``d3+'' for the first, second and third-up-to-infinity states, whilst the continuum is abbreviated as ``cont".}}
\label{fig:abs_coeff_gaas_mapi} 
\end{figure*}

\begin{table*}[!htb]
\caption{\label{table:gaas}GaAs best fitted parameters of the EBF model for temperatures in the range of $21$-$294$K. \blue{The error bars, presented in parenthesis, correspond to the uncertainties of the last digits, e.g., 27.8(13) means $27.8\pm1.3$, likewise, 0.91(3) is $0.91\pm0.03$.}}
\begin{indented}
\item[]\begin{center}
\begin{tabular}{@{}ccccccc}
\br
T(K)&A1&A2&$\sigma_{d}$ (meV)&$\sigma_{c}$ (meV)&$E_{g}$ (meV)&$E_{b}$ (meV)\\
\mr
21&27.8 (13) &0.91 (3) &2.11 (4) &2.34 (10) &1520 (0) &3.12(13)\\
90&28.94 (68)&0.88 (2)&2.31 (5)&2.43 (12)&1510 (0)&2.93 (6)\\
186&29.9 (14)&0.86 (3)&2.25 (6)& 2.42 (15)&1477 (21)&2.56 (12)\\
294&26 (12)&0.98 (46)&2.4 (10)&2.2 (16)&1425 (2)&2.08 (11)\\
  \br
\end{tabular}
\end{center}
\end{indented}
\end{table*}

\begin{table*}[!htb]
\caption{\label{table:mapbbr3}$\textrm{MAPbBr}_{3}$ best fitted parameters of the EBF model for temperatures in the range of $7$-$298$K}
\begin{indented}
\item[]\begin{center}
\begin{tabular}{@{}ccccccc}
\br
T(K)&A1&A2&$\sigma_{d}$ (meV)&$\sigma_{c}$ (meV)&$E_{g}$ (meV)&$E_{b}$ (meV)\\
\mr
7& 9.58 (18)& 0.72 (1)& 10.45 (4)& 7.90 (15)& 2285(1)&29.70(50)\\
20& 10.4 (3)& 0.68 (1)& 10.13 (7)& 7.37 (25)& 2287 (1) &27.41 (76)\\
40& 12.17 (52)& 0.65 (2)& 10.01 (9)& 6.96 (34)& 2292 (1)&24.01 (93)\\
60& 11.58 (31)& 0.68 (1)& 10.24 (6)& 7.27 (22)& 2303 (1)&25.64 (64)\\
80& 12.84 (35)& 0.69 (1)& 9.83 (6)& 6.57 (21)& 2307 (1)&23.13 (59)\\
100& 12.16 (25)& 0.73 (1)& 9.84 (6)& 6.55 (16)& 2314 (1)&24.07 (47) \\
120& 13.51 (44)& 0.69 (1)& 9.60 (9)& 5.91 (28)& 2315 (1)&22.20 (68)\\
140& 11.50 (31)& 0.74 (1)& 10.56 (8)& 6.44 (26)& 2326 (1)&26.46 (68)\\
160& 11.09 (21)& 0.77 (1)& 11.00 (6)& 6.83 (19)& 2332 (1)&27.40 (51)\\
180& 10.99 (21)& 0.78 (1)& 12.01 (7)& 7.69 (21)& 2340 (1)&28.28 (53)\\
200& 11.13 (28)& 0.81 (1)& 12.59 (1)& 7.99 (29)& 2345 (1)&27.46 (69)\\
220& 10.59 (23)& 0.78 (1)& 13.82 (11)& 9.02 (28)& 2353 (1)&28.80 (65)\\
240& 10.34 (30)& 0.83 (1)& 14.84 (14)& 10.13 (34)& 2359 (1)&29.17 (88)\\
260& 9.75 (20)& 0.86 (1)& 15.50 (12)& 10.57 (26)& 2365 (1)&30.16 (69)\\
280& 8.84 (23)& 0.91 (1)& 16.86 (13)& 12.44 (27)& 2374 (1)&33.02 (93)\\
298& 8.53 (11)& 0.89 (1)& 18.07 (4)& 13.60 (12)& 2381 (1)&33.28 (47)\\
  \br
\end{tabular}
\end{center}
\end{indented}
\end{table*}

\begin{figure*}[!htb]\centering % Using \begin{figure*} makes the figure take up the entire width of the page
\includegraphics[scale=0.29]{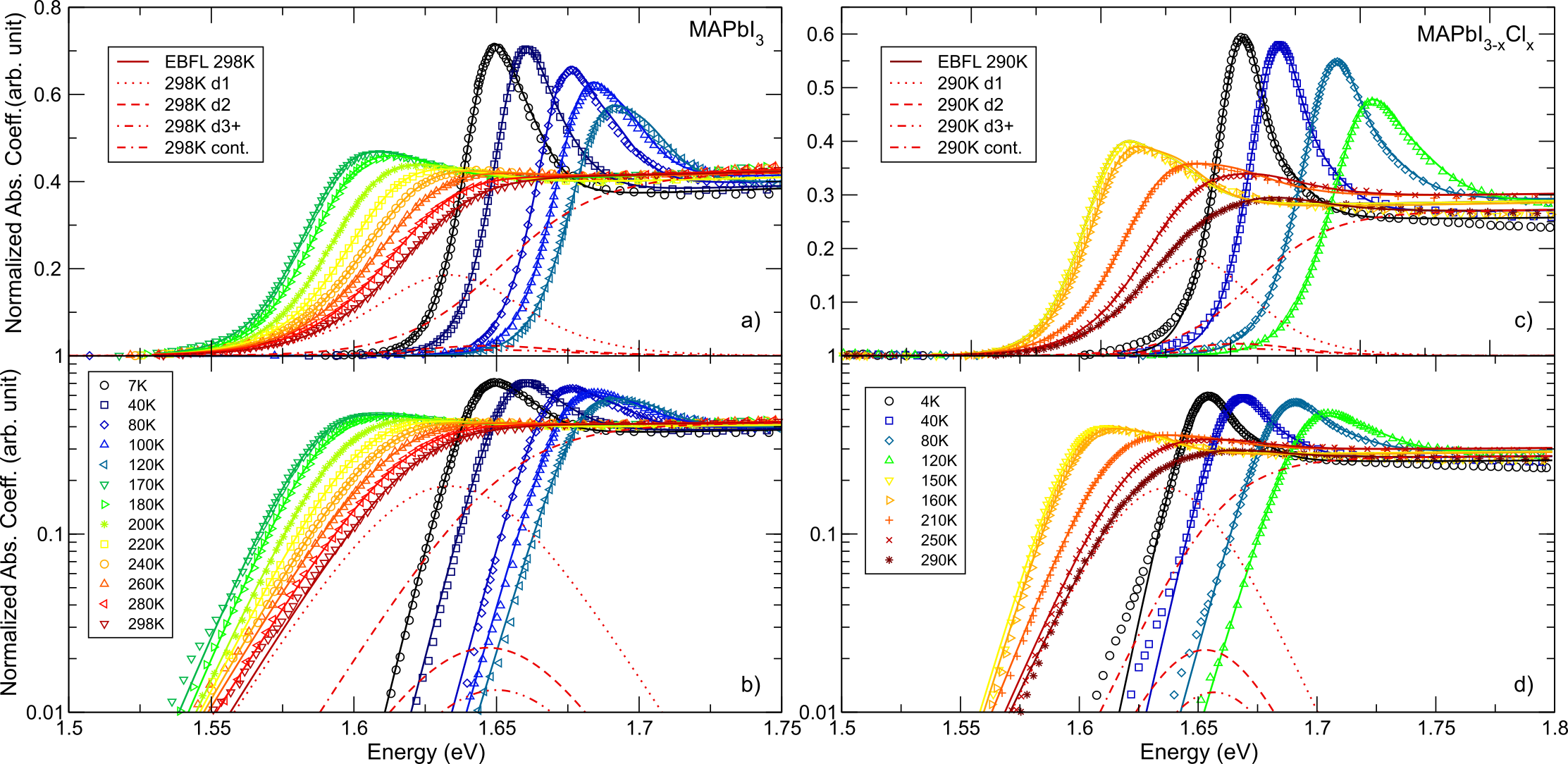}
\caption{Normalized absorption coefficient of $\textrm{MAPbI}_{3}$, extracted from Soufiani \cite{soufiani}, fitted with the EBF model  for several temperatures in linear (a) and logarithmic (b) scale. Figure (a) shows the contributions of the discrete peaks and the continuum, in dashed lines, for T=298K. Normalized absorption coefficient of $\textrm{MAPbI}_{3-x}\textrm{Cl}_{x}$, extracted from D’Innocenzo et al. \cite{dinn}, fitted with the EBF model for several temperatures in linear (c) and logarithmic scale (d). The contributions of the discrete and continuum parts of the spectra at T=290K are shown in red dashed lines, only, for viewing reasons. \blue{Here the excitonic peaks are denoted as ``d1'', ``d2'' and ``d3+'' for the first, second and third-up-to-infinity states, whilst the continuum is abbreviated as ``cont".}}
\label{fig:abs_coeff_mapi_mapcl}
\end{figure*}

\begin{table*}[!htb]
\caption{\label{table:mapbi3}$\textrm{MAPbI}_{3}$ best fitted parameters of the EBF model for temperatures in the range of $7$-$298$K}
\begin{indented}
\item[]\begin{center}
\begin{tabular}{@{}ccccccc}
\br
T(K)&A1&A2&$\sigma_{d}$ (meV)&$\sigma_{c}$ (meV)&$E_{g}$ (meV)&$E_{b}$ (meV)\\
\mr
7&7.39 (21)&0.68 (1)&6.62 (4)&4.93 (13)&1662 (1)&15.17 (38)\\
40&6.62 (29)&0.75 (2)&6.92 (5)&5.68 (18)&1675 (1)&16.85 (68)\\
80&6.25 (15)&0.81 (1)&7.23 (6)&4.83 (16)&1692 (1)&17.57 (42)\\
100&6.68 (17)&0.82 (1)&7.61 (7)&5.08 (17)&1696 (1)&16.32 (44)\\
120&6.47 (18)&0.83 (1)&8.11 (8)&5.28 (20)&1703 (1)&16.08 (46)\\
170&4.97 (14)&0.99 (2)&11.49 (8)&9.32 (19)&1613 (1)&17.86 (55)\\
180&5.37 (12)&0.96 (1)&11.52 (7)&9.27 (20)&1614 (1)&15.96 (37)\\
200&4.65 (12)&1.03 (2)&12.68 (8)&10.47 (20)&1624 (1)&18.34 (54)\\
220&4.84 (20)&1.01 (2)&14.04 (8)&12.56 (23)&1631 (1)&17.89 (80)\\
240&4.53 (37)&1.08 (5)&15.24 (9)&14.62 (24)&1639 (2)&19.3 (17)\\
260&4.61 (73)&1.08 (9)&16.19 (20)&15.47 (51)&1643 (5)&18.8 (32)\\
280&5.23 (81)&1.04 (10)&16.76 (12)&16.33 (61)&1644 (4)&15.5 (24)\\
298&4.24 (74)&1.22 (10)&16.70 (24)&15.45 (49)&1649 (3)&17.7 (15)\\
  \br
\end{tabular}
\end{center}
\end{indented}
\end{table*}

\begin{table*}[!htb]
\caption{\label{table:mapbi3cl3-x}$\textrm{MAPbI}_{x}\textrm{Cl}_{3-x}$ best fitted parameters of the EBF model for temperatures in the range of $4$-$290$K}
\begin{indented}
\item[]\begin{center}
\begin{tabular}{@{}ccccccc}
\br
T(K)&A1&A2&$\sigma_{d}$ (meV)&$\sigma_{c}$ (meV)&$E_{g}$ (meV)&$E_{b}$ (meV)\\
\mr
4&3.58 (11)&0.79(1) &6.85 (6)&5.24 (27)&1677 (1)&23.54 (67)\\
40&4.03 (15)&0.77(2) &7.32 (6)&6.10 (25)&1690 (1)&22.22 (76)\\
80&3.89 (7)&0.83(1) &8.76 (4)&6.98 (16)&1714 (1)&24.61 (44)\\
120&4.05 (7)&0.82(1) &9.63 (4)&7.21 (11)&1724 (1)&22.74 (39)\\
150&3.51 (8)&0.91(1) &9.76 (9)&6.18 (23)&1628 (1)&21.47 (56)\\
160&3.57 (9)&0.89(1) &10.02 (9)&6.29 (23)&1630 (1)&20.95 (55)\\
210&3.19 (6)&1.02(1) &12.87 (9)&9.22 (20)&1648 (1)&23.82 (58)\\
250&3.21 (7)&1.04(1) &14.22 (13)&10.51 (28)&1656 (1)&22.36 (62)\\
290&2.67 (6)&1.13(1) &15.17 (19)&10.96 (33)&1658 (1)&22.40 (68)\\
  \br
\end{tabular}
\end{center}
\end{indented}
\end{table*}

The absorption of direct GaAs is shown in figure  \ref{fig:abs_coeff_gaas_mapi}a,b in linear and logarithmic scale with an excellent agreement with the collected data. These figures shows the full absorption in solid lines, whilst the discrete and continuous contributions are plotted in dashed lines, for each temperature. Table \ref{table:gaas} summarize the best fitted parameters.

The absorption of tri-halide perovskites, $\textrm{MAPbBr}_{3}$, $\textrm{MAPbI}_{3}$ and $\textrm{MAPbI}_{3-x}\textrm{Cl}_{x}$ is presented as symbols in figures \ref{fig:abs_coeff_gaas_mapi}c,d, \ref{fig:abs_coeff_mapi_mapcl}a,b and \ref{fig:abs_coeff_mapi_mapcl}c,d, respectively. Here, the fitted absorption is presented in solid lines, while separated contributions are depicted in dashed lines. Our fits show that the three types of halide perovskites shape accurately the experimental curves. The best fitted parameters are summarized in tables \ref{table:mapbbr3}, \ref{table:mapbi3} and \ref{table:mapbi3cl3-x} for $\textrm{MAPbBr}_{3}$, $\textrm{MAPbI}_{3}$ and $\textrm{MAPbI}_{3-x}\textrm{Cl}_{x}$ respectively. 
% \begin{figure}
% \includegraphics[scale=0.3]{abs_coeff_gaas_mapbbr3_2.png}% Here is how to import EPS art
% \caption{\label{fig:abs_coeff_gaas_mapi} Absorption coefficient of GaAs for several temperatures fitted with the EBF model in linear (a) and logarithmic (b) scale. The inset of figure b shows all the contributions from the discrete peaks to the continuum at 294K. The absorption coefficient of $\textrm{MAPbBr}_{3}$ for several temperatures fitted with EBF model in linear scale is (c) and logarithmic scale (d). The inset of figure d shows the fitting procedure for the Soufiani Theory (GL) and the EBF with all of its contributions at 298K.}
% \end{figure}

In the case of GaAs, we observe a declining of the bandgap towards high temperatures. Whilst, in the case of perovskites, we report an increasing behavior of it. This difference is be described by the thermal expansion and electron-phonon interaction explained in detail in the next section. In addition, for $\textrm{MAPbI}_{3}$ and $\textrm{MAPbI}_{3-x}\textrm{Cl}_{x}$ we distinguish the well know phase transition from orthorombic to tetragonal phase marked at 140K with the notoriously kink of the bandgap. \magenta{However, in the case of the $\textrm{MAPbBr}_{3}$, the kink is less evident, but it can be inferred from other studies on the specific heat of tri-halide perovskites} \cite{onoda}.

\subsection{The Optical Bandgap vs T}
\begin{figure*}[ht] \centering
\includegraphics[scale=0.4]{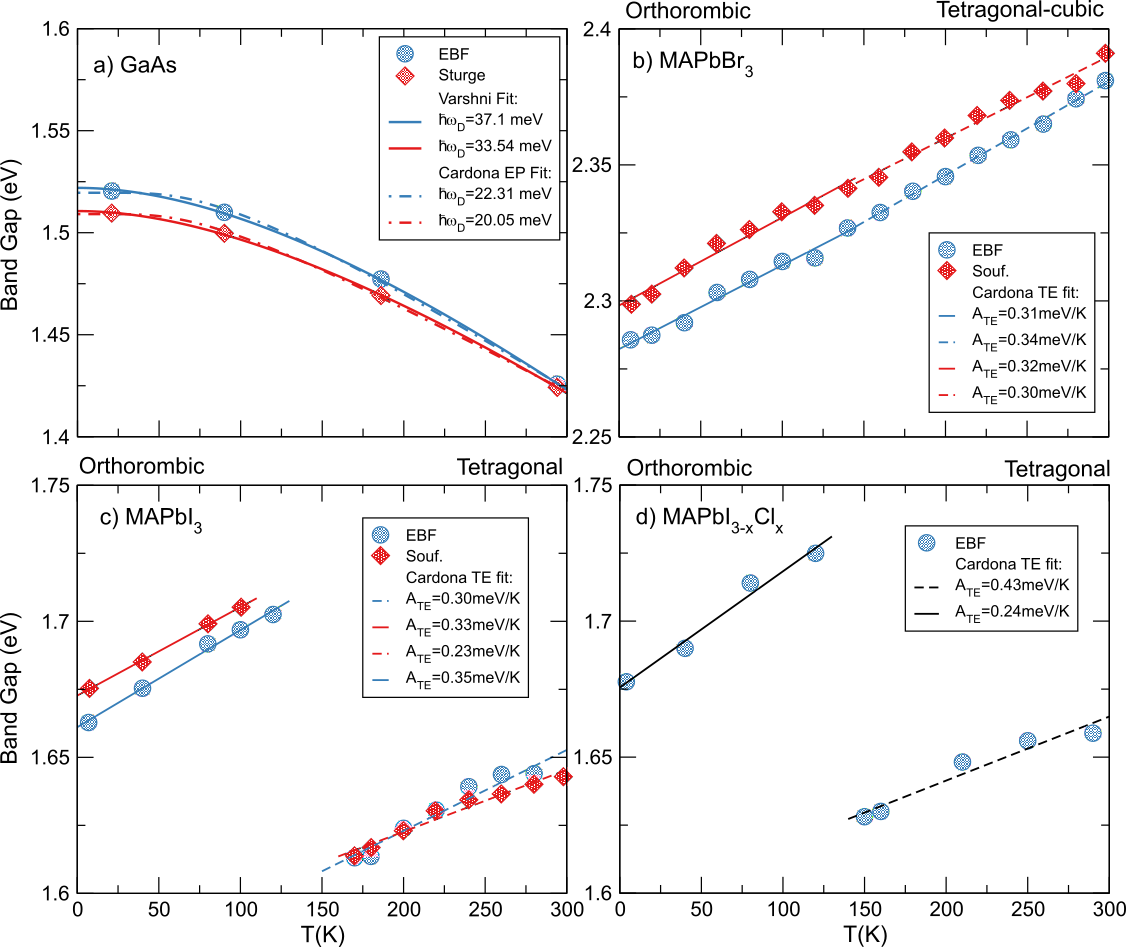}% Here is how to import EPS art
\caption{Bandgap energies for several temperatures, obtained by the EBF model, for GaAs (a), $\textrm{MAPbBr}_{3}$ (b), $\textrm{MAPbI}_{3}$ (c) and $\textrm{MAPbI}_{3-x}\textrm{Cl}_{x}$ (d) presented as blue marks. Figure (a) present the data collected from Sturge \cite{sturge} in red marks for comparison purposes. Here, the continuum and dashed lines show the fits using the Varshni model, eq. (\ref{varshni}), and Cardona's model, eq. (\ref{eqwei} with $A_{TE}=0$, respectively. Likewise, Figures (b) and (c) present the data collected from Soufiani et al. \cite{soufiani}, in red marks, for $\textrm{MAPbBr}_{3}$ and $\textrm{MAPbI}_{3}$ for comparison purposes. The figures (b), (c) and (d) show a set of continuum and dashed lines which represent the fit of Cardona's model, eq. (\ref{eqwei}) with $A_{EP}=0$, for the different perovskite phases, the orthorombic and tetragonal, respectively.}
\label{fig:bandgap}
\end{figure*}

\begin{table*}
\caption{\label{table:bandgap1} \blue{Best fitted parameters after} eq. (\ref{eqwei}) for GaAs and tri-halide perovskites at different temperatures. Here, the error bars are not shown due to their large values. However, it is remarkable that the LO phonon energy retrieved by this model is in agreement with the one calculated from the FWHM fit shown at table \ref{table:fwhm}. The fit of the set of data for Sturge \cite{sturge} and Soufiani \cite{soufiani} are labeled as the superscripts ($^{st}$) and ($^{s}$), respectively.}
\begin{indented}
\item[]\begin{center}
\begin{tabular}{@{}cccccc}
\br
Material & T (K) & $E_{0}$ (meV) & $A_{TE}$ (meV/K) & $A_{EP}$ (meV) & $\hbar\omega_{LO}$ (meV)\\
\mr
GaAs & 21-294 & 1617 & -0.12 & -94.78 & 35.74 \\
 &  & 1583$^{st}$ & -0.11$^{st}$ & -71.19$^{st}$ & 32.71$^{st}$ \\ \mr
$\textrm{MAPbBr}_{3}$ & 7-140 & 2283  & 0.31  & -1.85 & 17.97 \\
& & 2328$^{s}$ & 0.45$^{s}$ & -33.94$^{s}$ & 19.03$^{s}$ \\
& 160-298 & 2279& 0.39& -5.81& 17.29 \\
& & 2356$^{s}$ & 0.21$^{s}$ & -193.41$^{s}$ & 16.85$^{s}$ \\ \mr
$\textrm{MAPbI}_{3}$ & 7-120 & 1679 & 0.42 & -20.4 & 18.83 \\
& & 1674 $^{s}$ & 0.33 $^{s}$ & -1.54 $^{s}$ & 16.44 $^{s}$ \\ 
& 170-298 &  1711 & 9.76 & -698.95 & 12.31   \\
& & 1714$^{s}$ & 13.88 $^{s}$ & -805.9$^{s}$ & 9.95$^{s}$\\\mr
$\textrm{MAPbI}_{x}\textrm{Cl}_{3-x}$ & 4-120 & 1691 & 0.47 & -16.29 & 19.56 \\
& 150-290 & 1749 & 3.27 & -408 & 20.72\\
  \br
\end{tabular}
\end{center}
\end{indented}
\end{table*}

% \begin{ruledtabular}
% \begin{tabular}{cccccc}
% Material & T & $E_{0}$(meV) & $A_{te}$(meV/K) & $A_{ep}$(meV) & $\hbar\omega_{D}$(meV)\\
% \hline \hline
% GaAs & 21-294 & 1522 & 0.80 & - & 37.07 \\
%  &  & 1774$^{st}$ & 0.64$^{st}$ & - & 32.81$^{st}$ \\ \hline \hline
% $\textrm{MAPbBr}_{3}$ & 7-140K & 2283 & 0.31 & -1.85 & 17.97 \\
% & & 2328$^{s}$ & 0.45$^{s}$ & -33.94$^{s}$ & 19.03$^{s}$ \\
% & 160-298K & 2279 & 0.39 & -5.81 & 17.2 \\
% & & 2356$^{s}$ & 0.21$^{s}$ & -193.41$^{s}$ & 16.85$^{s}$ \\ \hline
% $\textrm{MAPbI}_{3}$ & 7-120K & 1679 & 0.426 & -20.4 & 18.83 \\
% & & 1674$^{s}$ & 0.33$^{s}$ & -1.53$^{s}$ & 16.44$^{s}$ \\ 
% & 170-298K & 1711 & 9.76 & -698.9 & 12.3 \\
% & & 1714$^{s}$ & 13.88$^{s}$ & -805.93$^{s}$ & 9.95$^{s}$ \\\hline
% $\textrm{MAPbI}_{x}\textrm{Cl}_{3-x}$ & 4-120K & 1691 & 0.47 & -16.29 & 19.56 \\
% & 150-290K & 1749 & 3.27 & -408 & 20.72 \\\hline
% \end{tabular}
% \end{ruledtabular}
% \end{table}
The evolution of the bandgap with temperature obtained after the fitting with the EBF model to GaAs and tri-halide perovskites absorption data are shown in figure \ref{fig:bandgap}. Here, we see that the results for GaAs are in close agreement with the values of Sturge \cite{sturge}, obtained by fitting the continuum part of the Elliot model at the band edge. In addition, other authors found similar results by measuring the reflectance for high purity and doped GaAs samples \cite{sell2,blakemore}. Likewise, Our results for $\textrm{MAPbBr}_{3}$ and $\textrm{MAPbI}_{3}$ are comparable with the values found at Soufiani et al \cite{soufiani}, obtained by fitting the pseudo Voigt of the Gaussian and the Lorentzian smearing. Additionally, these values are in agreement with other literature reports  \cite{galkowski,green,fabian,miyata}, confirming the good estimation of our results. Lastly, for $\textrm{MAPbI}_{3-x}\textrm{Cl}_{x}$, our results differs by $81$ meV, at low temperatures ($2-4$K), and by an amount of $46$ meV, at high temperatures ($190-200$K), with the report of Galkowski et al. \cite{galkowski}, using the magneto absorption technique. \blue{Nevertheless, this difference can be attributed to the sample composition due to the different methods used to synthesize $\textrm{MAPbI}_{3-x}\textrm{Cl}_{x}$ in D'Innocenzo et al. \cite{dinn}, and Galkowski et al. \cite{galkowski}.}

For the case of GaAs, the bandgap is reduced when temperature is increased, while in the case of tri-halide perovskites, the bandgap increments. The observed difference can be explained by the contributors of the bandgap evolution with T. These are the thermal expansion (TE) and the electron-phonon (EP) interaction \cite{francisco2019,wei}. For this reason, the evolution of the bandgap can be modeled as \cite{cardona2}:
\begin{equation}
 \Delta E_{g}(T)=\left[ \Delta E_{g}(T) \right]_{TE} + \left[ \Delta E_{g}(T) \right]_{EP}.  \label{deltaeg}
\end{equation}

The first term in the right hand of eq. (\ref{deltaeg}) is the effect of the bandgap caused by the contraction/expansion of the lattice as the response of the band-structure when an external hydrostatic pressure is applied  \cite{Lautenschlager,gopalan}. Whilst the second term is composed by the Debye-Waller and Fan terms. The former accounts for the interaction of electron with two phonons, whilst the latter describes the second order interaction of electron-phonon known as self energy \cite{gopalan,cardona2,gobel}. 

The thermal expansion contribution depends linearly with temperature \cite{cardona2}, i.e.,
\begin{equation}
    \Delta E_{g}(T) \mid_{TE} = -\alpha_{\nu} B_{0} \frac{dE_{g}}{dP} T, \label{thermal}
\end{equation}
%The evolution of the bandgap with the temperature  deals with two terms, the thermal expansion (TE) and the relinearization of the electron-phonon interaction (EP) which 
where $\alpha_{\nu}$ is the volumetric expansion coefficient, $B_{0}$ is the bulk modulus, and $dE_{g}/dP$ is the pressure behavior of the gap \cite{francisco2019,mannino}. The coefficients $\alpha_{\nu}$ and $B_0$ are positive for semiconductors such as Si\cite{okada}, Ge\cite{novikova}, GaAs \cite{novikovaGaAs} and halide perovskite families  \cite{Schueller,Rakita,sunalpha,wang,sunfang,Rodova,Zhang2017,wang2017}. Thus, implying that the sign of eq. (\ref{thermal}) depends exclusively on the pressure term, which is positive for semiconductors such as GaAs \cite{shakir} or Ge \cite{olguin,Lautenschlager}, and negative in cases like Si\cite{Lautenschlager} and tri-halide perovskites, $\textrm{MAPbI}_{3}$ \cite{francisco2018,francisco2019}, $\textrm{MAPbBr}_{3}$ \cite{kong}, $\textrm{MAPbCl}_{3}$  \cite{wang}.
%$\textrm{FAPbBr}_{3}$, $\textrm{CsPbBr}_{3}$  \cite{mannino}. 
%In the case of GaAs, the negative thermal expansion  this term is responsible for the bandgap reduction. Nevertheless, the pressure term is negative for families of halide perovskites as is found in the following reports, . This causes a positive thermal expansion for the bandgap evolution. Even though this could be thought as an explanation for the positive character, we are still left to understand the e-ph term

On the other hand, the renormalized bandgap due to the electron-phonon term is derived in works of Gopalan, Cardona, Lautenschlager and others \cite{gopalan,Lautenschlager,cardona2}. They deduce an expression where all the phonons modes of the $j^{th}$ branch, wave vector $q$ and energy $\omega_{jq}$ contribute to the shift and broadening of the electron state, $E_{nk}$, with wave vector $k$ at the $n^{th}$ electronic band \cite{francisco2019}, i.e.,
\begin{equation}
    \Delta E_{nk}(T)=\sum_{jq} \frac{\partial E_{nk}}{\partial n_{jq}} \left( n_{jq}(T) + \frac{1}{2} \right).
\end{equation}
\blue{Here $n_{jq}$ is the Bose-Einstein distribution, $n_{jq}=(e^{\beta\hbar\omega_{jq}}-1)^{-1}$ with $\beta=(k_{\beta}T)^{-1}$}, and the coefficients $\partial E_{nk}/\partial n_{jq}$ are the electron-phonon matrix elements. These elements can be taken as effective electron-phonon coefficients, $A_{i}$, at an average frequency $\omega_i$ which is deduced from the observation of the phonon DOS \cite{francisco2019,gobel,bhosale}. The resulting contribution of the EP interaction, known as Einstein oscillator, for the \blue{renormalization of the }bandgap is:
\begin{equation}
    \Delta E_{g}(T) \mid_{EP}=\sum_{i} A_{i} \left( n(\omega_i,T) +\frac{1}{2} \right). 
\end{equation}
\blue{Note that the $A_{i}$ coefficients are determined by the fitting of the experimental data.} However, this sum can collapse into a single term for cases such as GaAs and tri-halide perovskites due to the non presence of separated behaviors in the DOS\cite{francisco2019}, \blue{i.e.,
\begin{equation}
    \Delta E_{g}(T) \mid_{EP}= \frac{A_{1}}{2} \left( \frac{2}{e^{\hbar\omega/k_{\beta}T}-1}+1 \right). \label{electronphonon}
\end{equation}}
Nevertheless, there are families of semiconductors such as the cuprous halides where two oscillators are required for the acoustic and optical branches \cite{gobel,serrano}. \blue{By replacing eqs. (\ref{thermal}) and  (\ref{electronphonon}) in eq. (\ref{deltaeg}), we can write the full expression for the description of the bandgap proposed by Cardona and co-workers,}
\begin{equation}
    \Delta E_{g}(T) = E_{g_{0}}+A_{TE}T+A_{EP} \left(\frac{2}{e^{\hbar\omega/k_{\beta}T}-1}+1 \right).
    \label{eqwei}
\end{equation}
Here, $E_{g_{0}}$ is the unrenormalized bandgap, and the coefficients are $A_{TE}=-\alpha_{\nu}B_{0}\frac{dE_{g}}{dP}$ and $A_{EP}=A_{1}/2$. \blue{$\omega$ is the average optical phonon frequency  ($\omega_{LO}$) which is also called Debye frequency ($\omega_D$) due to its theoretical closeness \cite{passler}}. 

In previous reports, we have found that the EP contribution of GaAs are around 79$\%$ \cite{lourenzo}, 67$\%$ \cite{hennel}, and 56$\%$ \cite{biernacki}. The reason why EP is larger than the TE for this material may be due to the small volume deformation potential as it has been shown for Si \cite{cardona2}. However, this behavior is not repeated for the family of tri-halide perovskites in which the EP interaction contributes only up to 40$\%$ \cite{francisco2019,wei,yu2011}. 
%In the case of silicon, a contribution of  in Street et al. \cite{street} for Si (98$\%$ of contribution) due to the small lattice constant and small TE contribution. 
%What is more, this e-ph term tends to be negative which could not explain the positive tendency of the perovskite's bandgap. Eventhough this balance between thermal expansion and e-ph term its a current problem. It is worth it to mention that the composition of the e-ph interaction are the Fan (self energy)\cite{fan} and Debye-Waller\cite{debyewaller,debyewaller2} as stated in , the bandgap evolution evolution for e-ph is:
%begin{equation}
%    \Delta E_{nk}(T)=\sum_{jq} \frac{\partial E_{nk}}{\partial n_{jq}} \left( n_{jq}(T) + \frac{1}{2} \right).
%\end{equation}
%Since both aforementioned contributions play a fundamental rol for the bandgap evolution. The theoretical equation of simple summatory is:
%\begin{equation}
%    \Delta E_{g} = E_{g_{0}}+A_{Te}T+A_{EP}(\frac{2}{1+Exp[\hbar\omega/k_{\beta}T]}+1)
%    \label{eqwei}
%\end{equation}
%Where the coefficients are in the units of meV/K and meV for the TE and EP respectively.

Eq. (\ref{eqwei}) was used to fit the bandgap evolution with temperature of $\textrm{GaAs}$, $\textrm{MAPbBr}_{3}$, $\textrm{MAPbI}_{3}$ and $\textrm{MAPbI}_{3-x}\textrm{Cl}_{x}$. \blue{During the implementation of the fit, our analysis encounter problems regarding the uncertainties of the parameters. For instance, the linear behavior of the perovskites overshadow the electron phonon term producing large error bars for the $A_{EP}$ term. Likewise, for GaAs, the detriment of the exponential opaque the linear behavior of the thermal expansion. Notwithstanding, the best fitted parameters using eq. (\ref{eqwei}) are presented in table \ref{table:bandgap1} for comparison purposes only.} There, we can see a good estimation of the LO phonon energy  comparable to reports found in \cite{yull,Landolt-Bornstein} and \cite{yang2017,leguy,gold} for GaAs and the \blue{tri-halide perovskites}, respectively. 

Despite the previous procedure, we have used a thermal expansion independent version of the Cardona's model \cite{cardona2}, i.e., eq. (\ref{eqwei}) with $A_{TE}=0$. This is then applied to fit the bandgap evolution of GaAs. Likewise, an independent electron-phonon version of Cardona's model, i.e. eq. (\ref{eqwei}) with $A_{EP}=0$, is used for the case of tri-halide perovskites. These procedures are based on the behavior of the experimental data who obeys separated trends. In addition, this choose is reinforced by previous studies on these materials \cite{soufiani,lourenzo}. Additionally, we have used the Varshni's model \cite{varshni}, that shapes the bandgap empirically, by taking the form of the e-ph interaction, only for the case of GaAs. Varshni's model reads as:
\begin{equation}
    E_{g}(T)=E_{0}+\frac{a_{B} T^2}{T+\hbar\omega_{D}/k_{\beta}}, \label{varshni}
\end{equation}
\blue{where $E_{0}$ is the bandgap at 0K, $a_{B}$ is the amplitude and $\omega_{D}$ is the Debye frequency.} The results of these analyses are collected in table \ref{table:bandgap2}, and presented in figure \ref{fig:bandgap} as continuum and dashed blue lines. In the case of tri-halide perovskites, these lines denote the orthorombic and tetragonal phases, respectively. Whereas in the case of GaAs, they denote the difference between the Varshni and Cardona's model. 

In addition, we have fitted the data of the bandgap extracted from Sturge \cite{sturge}, and Soufiani et al. \cite{soufiani}. These lines are colored as red dashed and continuum lines. They show a good agreement with our results from the EBF model. Remarkably, the Debye energy obtained with the Varshni's model presented in table \ref{table:bandgap2} for GaAs is in agreement with other experimental values of $28$ meV \cite{passler2}, $28.87$ meV \cite{yull,Landolt-Bornstein}, $36.1$ meV \cite{nsm}, and $36.8$ meV \cite{rudin}, and with theoretical calculations of phonon DOS found in Lawler et al. \cite{lawler}. However, when these values are compared with the independent version of the Cardona's model, \blue{(eq. (\ref{eqwei}) with $A_{TE}=0$)}, we appreciate a difference of $15$ meV that can be attributed to the fitting goodness.

%a Here we see similar values for the parameters $E_0$, $A_{te}$ and $\hbar\omega_D$ for the EBF data and the literature sets for GaAs and halide perovskites. For the latter, the difference of $A_{ep}$ is clearly seen in figure \ref{fig:bandgap}b where the larger curvature, only shown for Soufiani data, is produced by a larger EP term. 
%These coefficients are are commonly shared for the tri-halide family and share a similar behavior to the ones found on Wei et al. \cite{wei}. On the other hand, since the amount of parameters is 4, we suffer the penalty of equal number of parameters and data points for the fit procedure on GaAs. For this reason we have used the empirical models for the description of the bandgap, where both contributions are included, for instance, we are using the Varshni model \cite{varshni}. However, many empirical models has been developed since then, such as the Passler \cite{passler}, Viña \cite{vina} and the two-oscillator model \cite{gobel,yu2011}. In the Varshni formualtion, the evolution of the bandgap is:
%where $E_g(0)$ is the bandgap at T=0 and $\hbar\omega_{D}$ is the phonon frequency mode which in this case is called the Debye temperature.

\begin{table*}
\caption{\label{table:bandgap2}Parameters of the bandgap evolution versus temperature of GaAs by using the Varshni model, eq. (\ref{varshni}), and the Cardona's model, eq. (\ref{eqwei}) with $A_{TE}=0$. In the case of tri-halide perovskites, only the linear thermal expansion contributes to eq. (\ref{eqwei}). Here, the fits of the set of data for Sturge \cite{sturge} and Soufiani \cite{soufiani} are labeled as the superscripts ($^{st}$) and ($^{s}$), respectively.}
\begin{indented}
\item[]\begin{center}
\begin{tabular}{@{}cccccc}
\br
Material & T (K) & \centre{4}{Model} \\ \mr
& & \centre{4}{Varshni et al. \cite{varshni}}  \\
\ns
 &  &  \crule{4} \\
& &  & $E_{0}$ (meV) & $a_{B}$ (meV/K) & $\hbar\omega_{D}$ (meV)\\
GaAs & 21-294 &  \crule{4} \\ 
&  &  & 1522 (1.09) & 0.80 (0.14)  & 37.08 (11.08) \\
&  &  & 1511 (0.73)$^{st}$ & 0.68 (0.08)$^{st}$  & 33.54 (7.11)$^{st}$ \\\mr
& & \centre{4}{Cardona et al. \cite{cardona}} \\
\ns
 &  &  \crule{4} \\
 & & $E_{g_{0}}$ (meV) & $A_{TE}$ (meV/K) & $A_{EP}$ (meV) & $\hbar\omega_{LO}$ (meV)\\
GaAs &  21-294 &  \crule{4} \\
&  & 1585 (15.33)  & 0 &  -65.9 (16.5) & 22.31 (3.79) \\
 &  & 1560 (9.19)$^{st}$ & 0 & -51.02 (10.03)$^{st}$  & 20.05 (2.79)$^{st}$  \\ \mr
& 7-140 &  2282 (1.52) & 0.31 (0.02)  & 0 & - \\
$\textrm{MAPbBr}_{3}$  & & 2298 (1.68)$^{s}$ & 0.32 (0.02)$^{s}$ & 0 & - \\
& 160-298 & 2277 (1.94) & 0.34 (0.008) & 0 & - \\
& & 2299 (4.39)$^{s}$ & 0.30 (0.02)$^{s}$ & 0 & - \\ \mr
 & 7-120 & 1661 (1.33) & 0.35 (0.02) & 0 & - \\
$\textrm{MAPbI}_{3}$& & 1673 (0.54)$^{s}$ & 0.33 (0.008)$^{s}$ & 0 & - \\ 
& 170-298 & 1564 (5.91)  & 0.30 (0.03) & 0 & -  \\
& & 1577 (4.08) $^{s}$ &  0.23 (0.02)$^{s}$ & 0 & - \\\mr
$\textrm{MAPbI}_{x}\textrm{Cl}_{3-x}$ & 4-120 & 1676 (3.29) & 0.43 (0.04) & 0 & - \\
 &  150-290 & 1594 (7.39) & 0.24 (0.03) & 0 & - \\
  \br
\end{tabular}
\end{center}
\end{indented}
\end{table*}

%\textcolor{red}{We may not know the proportion between the Frolich coupling constant and the EP term in the bandgap}
%\textcolor{red}{which can be added by the addition of the octahedral tilting \cite{mannino}}
\subsection{The Full Width Half Maximum (FWHM) vs T}
\begin{figure*}[ht] \centering
\includegraphics[scale=0.4]{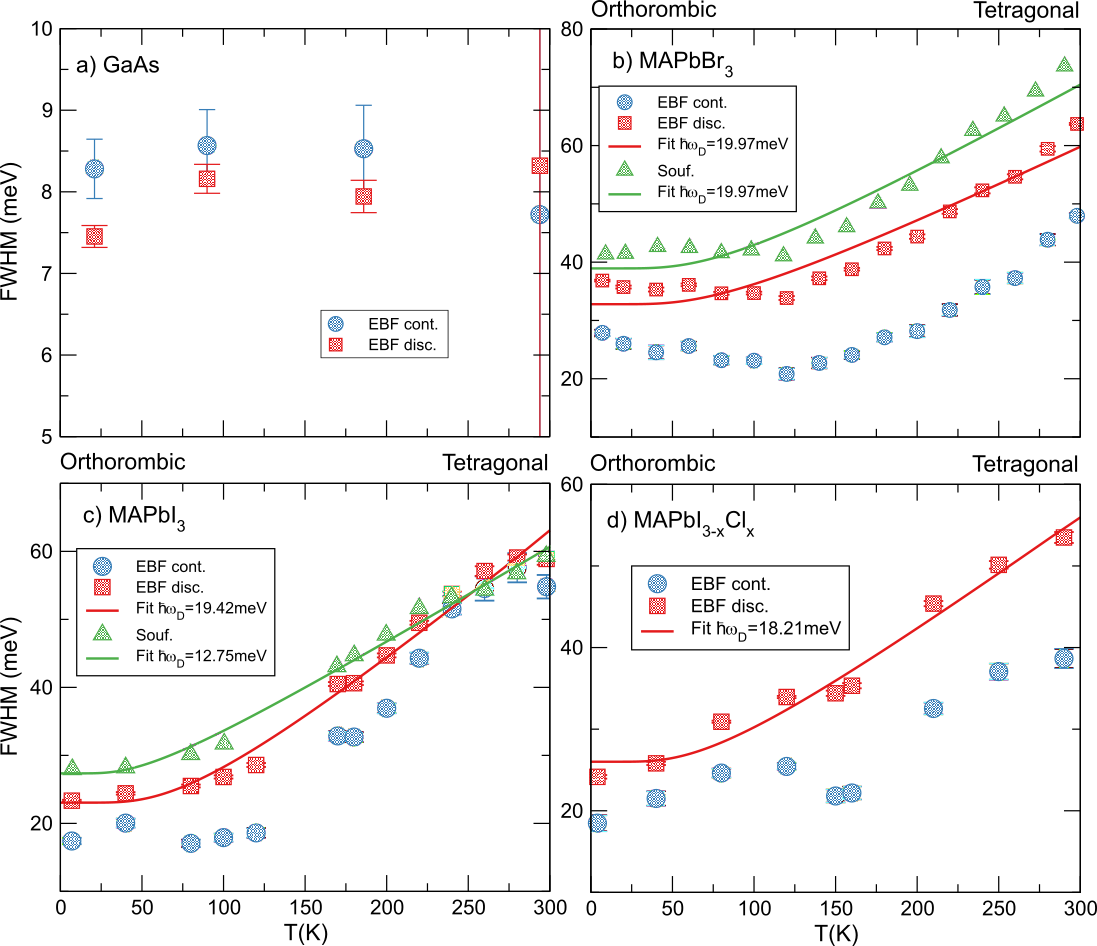}% Here is how to import EPS art
\caption{FWHM retrieved by the EBF model for the discrete and continuum contributions marked as blue and red points, respectively, for GaAs (a), $\textrm{MAPbBr}_{3}$ (b), $\textrm{MAPbI}_{3}$ (c) and  $\textrm{MAPbI}_{3-x}\textrm{Cl}_{x}$ (d). For comparison purposes, we have extracted the data from Soufiani \cite{soufiani} and plotted in green triangles in figures (b) and (c). Eq. (\ref{linewidth}) with $\Gamma_{AC}=0$ is used to fit the excitonic linewidth of the discrete peak for the tri-halide perovskites in figures (b), (c) and (d). This is presented as red and green lines for the EBF and Soufiani data, respectively.}
\label{fig:fwhm} 
\end{figure*}

\blue{The full width half maximum represents the broadening of the measurement associated to thermal or uncertainty contributions. This is presented in figure \ref{fig:fwhm} for the GaAs and tri-halide perovskites. In our case, the FWHM is only modeled by thermal effects that are shaped by the EBF model. This were calculated by using $4\sigma_{n}\textrm{arccosh}(\sqrt{2})$ which is a consequence of the potential fluctuations of eq. (\ref{w}). In the analysis, we found  contributions associated with the discrete, $\sigma_{d}$, and continuum, $\sigma_{c}$, parts of the spectra. On the one hand, the discrete case corresponds to the broadening of the excitonic peak due to the aforementioned thermal contribution. Whereas,} in the continuum case, this energy corresponds to the Urbach energy associated with the disorder induced localized states.

\blue{Figure \ref{fig:fwhm}a shows a horizontal behavior of the discrete and continuum linewidths for GaAs. Here, at $294$K, the values show large error bars due to non resolved excitonic peak observed in the absorption spectrum at this temperature. For the case of $\textrm{MAPbBr}_3$, figure \ref{fig:fwhm}b shows a horizontal trend for the orthorombic phase, while an increasing behavior is observed for the tetragonal phase. On the other hand, figures \ref{fig:fwhm}c and d show a raising trend of the discrete linewidth versus temperature for $\textrm{MAPbI}_3$ and $\textrm{MAPbI}_{3-x}\textrm{Cl}_{x}$, respectively.}

The linewidth values for GaAs, at high temperatures ($200-300$K),  are in agreement with previous reports \cite{golap,qiang}. Likewise, for the case of $\textrm{MAPbBr}_{3}$ and $\textrm{MAPbI}_{3}$, the linewidths are comparable to the values reported at \cite{soufiani,adam,fang,diab}, \blue{for the whole studied temperature range of 0-300K.}  Lastly, the results of $\textrm{MAPbI}_{3-x}\textrm{Cl}_{x}$, at room temperature, \blue{show a difference of $50$ meV and $30$ meV when compared with the PL reports of Wehrenfennig et al. \cite{christian}, and, Wu et al. \cite{wukw}, respectively.} It is remarkable to state that the values obtained from the linewidth evolution of the EBF model are similar to other studies of the families of tri-halide perovskites based on cesium, Cs, \cite{wei,saran} or formamidinium, FA, \cite{adam}. 

The thermal evolution of the discrete excitonic linewidth can be modeled in terms of the interaction of the exciton and phonons which is modeled with the electron-phonon interaction. The theory describing the evolution of the linewidth was developed initially by Toyozawa \cite{toyozawa2}, Segall and Mahan \cite{segall}. They divided the contributions in the interactions between electrons and acoustical phonons, and among electrons with optical phonons. In the former case, the linewidth is calculated from the deformation potential and piezoelectric interactions \cite{rudin} which gives a linear evolution with temperature. In the latter, the interaction with LO phonons is modeled by the Fr$\ddot{\textrm{o}}$lich interaction which produces a linewidth dependent on the Bose-Einstein distribution, i.e., proportional to $(e^{\hbar\omega_{LO}/k_{\beta}T}-1)^{-1}$ \cite{rudin}. As a consequence, the final form of the linewidth can be written as:
\begin{equation}
  \Gamma(T)=\Gamma_{0}+\Gamma_{AC}T+\Gamma_{LO} \left( \frac{1}{e^{\hbar\omega_{LO}/k_{\beta}T}-1} \right). \label{linewidth}
\end{equation}
Here $\Gamma_{0}$ corresponds to linewidth at 0K, whilst $\Gamma_{AC}$ and $\Gamma_{LO}$  are the proportional constants related to acoustical and optical phonon terms. Additionally, $\hbar\omega_{LO}$ is the LO phonon energy. 

For our analyzed materials, we are considering eq. (\ref{linewidth}) to be independent of the acoustical contribution, i.e., $\Gamma_{AC}=0$. This decision is based on previous studies of GaAs \cite{rudin} and tri-halide perovskites \cite{adam,bernhard}. And, in the case of perovskites this is reinforced by their their polar nature. The retrieved values of this procedure are presented in table \ref{table:fwhm}. There, we see that for tri-halide perovskites, the fitted LO phonon energies after the EBF model are $19.97$ meV, $19.42$ meV and $18.21$ meV for $\textrm{MAPbBr}_{3}$, $\textrm{MAPbI}_{3}$ and $\textrm{MAPbI}_{3-x}\textrm{Cl}_{x}$, respectively. In the case of $\textrm{MAPbBr}_{3}$ and $\textrm{MAPbI}_{3}$, we have applied eq. (\ref{linewidth}) to the FWHM obtained in Soufiani's work, for comparison purposes. The fit, presented in green lines in figures \ref{fig:fwhm}b and \ref{fig:fwhm}c, show a similar trend when compared to our results. In addition, the obtained LO phonon energy are within the range of $0-20$ meV of the phonon DOS calculated with first principles  \cite{yang2017,leguy,gold}. 

For the case of GaAs, the flatness of its linewidth make the fit imprecise, causing large uncertainties of the LO phonon energy. This behavior can be attributed to the lack of points of the absorption data (see figures \ref{fig:abs_coeff_gaas_mapi}a,b) around the exciton peak. \blue{Nevertheless, more tests on GaAs and other members of the group III-IV semiconductors need to be analyzed in order to confirm this trend.} 

Lastly, It is remarkable that the computed LO phonon energy, here obtained with the linewidth for tri-halide perovskites, is an stepping stone for the theoretical calculation of the binding energy of excitons in polar semiconductors as we will present in the next section.
%If $\sigma_{d}$ is large, then the exciton is more diffuse in energy as is presented in figure \ref{fig:abs_coeff_gaas_mapi}a,c and \ref{fig:abs_coeff_mapi_mapcl}a,c for room temperature.
\begin{table*}[ht]
\caption{\label{table:fwhm} Parameters of the linewidth evolution with the temperature of the EBF model by fitting eq. (\ref{linewidth}) for tri-halide perovskites. Note that in all cases the acoustic term $\Gamma_{AC}$ is set to zero due to the large Fr$\ddot{\textrm{o}}$lich interaction of the optical phonons for these semiconductors. Eq. \ref{linewidth} is applied to the data of Sturge \cite{sturge} and Soufiani \cite{soufiani} as well, and the results are labeled with the superscripts ($^{st}$) and ($^{s}$), respectively.}
\begin{indented}
\item[]\begin{center}
\begin{tabular}{@{}cccc}
\br
Material & $\Gamma_{0}$(meV) & $\Gamma_{LO}$(meV) & $\hbar\omega_{LO}$(meV) \\
\mr
%GaAs & 0.072 (0.068 \cite{sendner}) & 0.077\cite{dubey} & 0.865 \cite{dubey} & 10.9\cite{dubey} & 12.9\cite{dubey} & 37.08 \\
%InAs & 0.052 & 0.026 & 0.026\\
%CdTe & 0.29 & 0.088 & 0.92\\
%GaAs &7.60 (0.25)  & 2.47e-03 (1.4e-03) & 0 & - \\
$\textrm{MAPbBr}_{3}$ & 32.78 (1.60)   & 31.47 (20.4) & 19.97 (9.04)\\
 & 38.93 (1.56)$^s$   & 36.73 (20.02)$^s$  & 19.97 (7.57)$^s$ \\
$\textrm{MAPbI}_{3}$  & 23.05 (1.28)  &  44.5 (14.19) & 19.42 (4.42)\\
 & 27.36 (1.04)$^s$   & 21.25 (7.61)$^s$ & 12.76 (3.70)$^s$\\
$\textrm{MAPbI}_{3-x}\textrm{Cl}_{x}$ & 26.02 (14.20)  & 30.7 (15.65) & 18.21 (6.65)\\
  \br
\end{tabular}
\end{center}
\end{indented}
\end{table*}
%Here we see the phase transition associated with perovskites. is clearly visible for the $\textrm{MAPbI}_{3}$ and  $\textrm{MAPbI}_{3-x}\textrm{Cl}_{x}$. While in the case of  $\textrm{MAPbBr}_{3}$, the difference is embedded in the tendency of the curve who changes from a flat behavior to a linear growth.

\subsection{Exciton Binding Energy}

%For instance, this interaction is significant for perovskites according to the Fr$\ddot{\textrm{o}}$lich coupling constant detailed in table \ref{table:frolich}, extracted from \cite{baranowski}.

%There we can see that in cases where e-ph interaction is negligible such as in GaAs or InAs, the effective electron mass does not change, while for perovskites, the effective electron mass change due to the cloud of phonons surrounding it.

\begin{table*}
\caption{\label{table:frolich} Fr$\ddot{\textrm{o}}$lich coupling constant ($\alpha$) of GaAs and tri-halide perovskites computed with eq. (\ref{eqfrolich}). The values used for the calculation were extracted from previous reports \cite{soufiani,dubey}. In the case of $\textrm{MAPbI}_{3-x}\textrm{Cl}_{x}$,  the values denoted as ($^*$) are an average of the parameters for $\textrm{MAPbI}_{3}$ \cite{soufiani} and $\textrm{MAPbCl}_{3}$\cite{sendner}. In addition, we have used the LO phonon energies retrieved by the Varshni equation for GaAs, and the FWHM fit for tri-halide perovskites. Lastly, previous values of $\alpha$ are written in parenthesis for comparison purposes. Note that the effective masses, $m_{e}$ and $m_{h}$, are in terms of the unit mass of electron. $\epsilon_{\infty}$ and $\epsilon_0$ are in units of the vacuum permittivity.}
\begin{indented}
\item[]\begin{center}
\begin{tabular}{@{}ccccccc}
\br
Material & $\alpha$ & $m_{e}$ & $m_{h}$ & $\epsilon_{\infty}$ & $\epsilon_0$ & $\hbar\omega_{LO}$ (meV) \\
\mr
GaAs & 0.072 (0.068 \cite{sendner}) & 0.077\cite{dubey} & 0.865 \cite{dubey} & 10.9\cite{dubey} & 12.9\cite{dubey} & 37.08 \\
%InAs & 0.052 & 0.026 & 0.026\\
%CdTe & 0.29 & 0.088 & 0.92\\
$\textrm{MAPbBr}_{3}$ & 1.90 (1.69\cite{sendner}) & 0.29 \cite{soufiani} & 0.31\cite{soufiani} & 4.4\cite{soufiani} & 25.5\cite{soufiani} & 19.97 \\
$\textrm{MAPbI}_{3}$ & 1.41 (1.72\cite{sendner})& 0.23\cite{soufiani} & 0.29\cite{soufiani} & 5\cite{soufiani} & 19.6\cite{soufiani} & 19.42 \\
$\textrm{MAPbI}_{3-x}\textrm{Cl}_{x}$ & 1.94 (1.95\cite{sendner})& 0.32$^{*}$ & 0.29$^{*}$ & 4.5$^{*}$ & 24.7$^{*}$ & 18.21 \\
  \br
\end{tabular}
\end{center}
\end{indented}
\end{table*}

\begin{figure*}[ht] \centering
\includegraphics[scale=0.4]{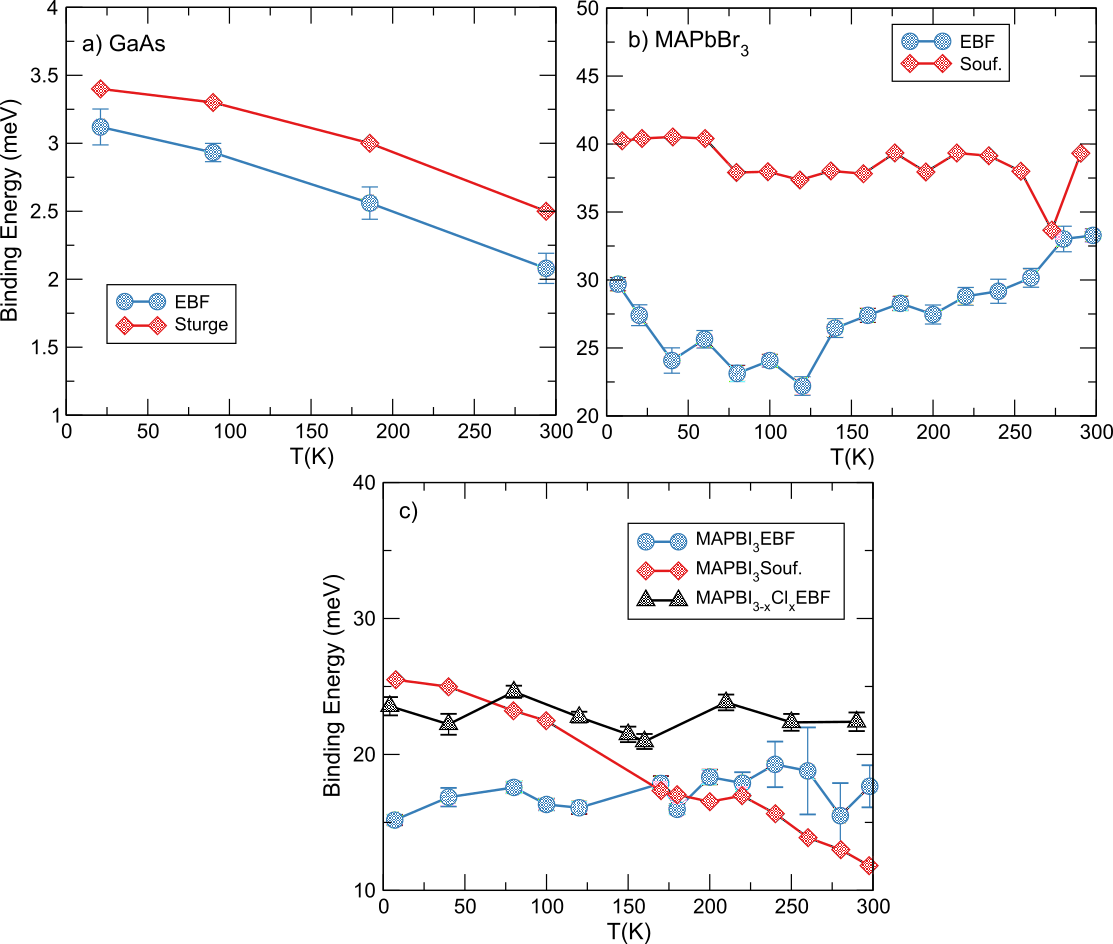}% Here is how to import EPS art
\caption{Binding energy of the first excitonic state versus temperature retrieved by fitting the EBF model in optical absorption data of GaAs (a), $\textrm{MAPbBr}_{3}$ (b), and $\textrm{MAPbI}_{3}$-$\textrm{MAPbI}_{3-x}\textrm{Cl}_{x}$ (c). The data of figure (a) is compared with the results extracted from Sturge \cite{sturge}, while in figures (b) and (c) these are compared with Soufiani values \cite{soufiani}. We note that the binding energies in figures (b) and (c), collected with the EBF model, show oscillations with average values of $27.52$ meV, $17.17$ meV and $22.68$ meV, for $\textrm{MAPbBr}_{3}$, $\textrm{MAPbI}_{3}$ and $\textrm{MAPbI}_{3-x}\textrm{Cl}_{x}$, respectively.}
\label{fig:bindingenergy} 
\end{figure*}

The results of the binding energy versus temperature for the 1s state of the exciton can be seen in figures \ref{fig:bindingenergy} for GaAs and tri-halide perovskites. Here, GaAs shows a decreasing behavior, while tri-halide perovskites show small fluctuations around an average energy value. In the former, the collected values are in the interval of $2$-$3.1$ eV. These values share an small difference with the values reported at Sturge et al., as is seen in figure \ref{fig:bindingenergy}a. However, other studies report values of $4.2$ meV \cite{nam,fehrenbach,sestu2}. \blue{This difference is attributed to the models for extracting the binding energy. For instance, Sestu et al. uses the sum rule on the absorption, Nam et al. retrieves it from PL and reflection measurements, and Fehrenbach et al. locates empirically the excitonic peaks from absorption measurements at low temperature.}

%Here the constant value of the binding energy is reflected deeply in $\textrm{MAPbI}_{3}$ and $\textrm{MAPbI}_{x}\textrm{Cl}_{3-x}$, while the values are more spread in the case of $\textrm{MAPbBr}_{3}$
In the case of tri-halide perovskites, we have values ranging from $22$-$33$ meV, $15$-$19$ meV and $21$-$25$ meV for $\textrm{MAPbBr}_{3}$, $\textrm{MAPbI}_{3}$ and $\textrm{MAPbI}_{x}\textrm{Cl}_{3-x}$, respectively. These values are compared with the results reported in Soufiani et al. \cite{soufiani}, plotted as red diamonds in figure \ref{fig:bindingenergy}b and c, where we see fluctuations of the binding energy for $\textrm{MAPbBr}_{3}$, whilst a detrimental behavior is shown for $\textrm{MAPbI}_{3}$.

The aforementioned binding energies are compared with other values from literature summarized in table \ref{table:binding_energy}. There, we observe that our obtained binding energies for  $\textrm{MAPbBr}_{3}$ are in agreement with temperature dependent PL experiments \cite{niesner,kunugita} and high field magnetoabsorption \cite{galkowski}. Likewise, for the case of $\textrm{MAPbI}_{3}$, our results are comparable with values obtained with absorption \cite{even}, high field magnetoabsorption \cite{miyata,galkowski} and PL \cite{sun}. However, for the case of $\textrm{MAPbI}_{x}\textrm{Cl}_{3-x}$, we do not have a coincidence with other reports despite being in the same order of magnitude. This can be attributed to the different sample preparations. However, our results are aligned with the theoretical prediction of the binding energy, whose value should be in the intermediate of iodide and chloride perovskites. In addition, according to our fitting results of bandgap and binding energy, the concentration of the mixed halide perovskite is closer to $\textrm{MAPbI}_{3}$, i.e., $x\sim3$. 

The binding energies in table \ref{table:binding_energy} show a large difference between several reports on the tri-halide perovskites. \blue{On the one hand, this might be attributed to differences in deposition conditions and, in particular, to degradation \cite{tejada}. On the other hand, it could be} attributed to the different techniques used \blue{to obtain the binding energy}, for instance we have absorption, photoluminescence and magnetoabsorption. Each of these experiences have different procedures to extract the binding energy. For example, in absorption, an Elliott fit is performed, in PL, the Arrhenius formula is employed, whilst in the case of magnetoabsorption, a modified Elliott equation with a magnetic potential is used \cite{hirasawa}. These models for excitons are based on the hydrogen atom which does not accurately describe the case of perovskite systems, since the different binding energies detected in experiments at low and high magnetic fields indicate a large presence of electron-phonon interaction \cite{baranowski}. For instance, the formation of polarons could explain the enhancement of the carrier effective mass for low magnetic fields while for high magnetic fields, the carrier-lattice interaction is decoupled \cite{baranowski}.
%Ideas that are based on the Hydrogen atom, which has a characteristic close difference between the optical ($\epsilon_{\infty}$) and static ($\epsilon_{s}$) dielectric  constant. 

Polarons in perovskites have been helpful when explaining the low mobility of carriers \cite{sendner}, the slow carrier cooling \cite{frost}, and their structural softness \cite{miyatak}. These polarons are formed by coupling the electronic states to the longitudinal optical (LO) phonons, so the electron is dressed by the lattice polarization with an effective mass, in the week regime, equivalent to $m^{*}_{e,h}=m_{e,h}(1+\alpha/6)$). Here, the Fr$\ddot{\textrm{o}}$lich coupling constant $\alpha$ is defined as \cite{Frohlic,baranowski}:
\begin{equation}
    \alpha=\frac{e^2}{\hbar}\left( \frac{1}{\epsilon_{\infty}}-\frac{1}{\epsilon_{s}} \right)\sqrt{\frac{m_{e,h}}{2\hbar\omega_{LO}}}. \label{eqfrolich}
\end{equation}

In eq. (\ref{eqfrolich}), $\epsilon_{\infty}$ and $\epsilon_{s}$ are the optical and static dielectric constant, respectively. And, $\hbar\omega_{LO}$ is the LO phonon energy. The combination of a large value of the effective dielectric screening ($1/\epsilon^{*}=1/\epsilon_{\infty}-1/\epsilon_{s}$) \cite{brivio} and a low phonon energies result in a high Fr$\ddot{\textrm{o}}$lich coupling constant as you can see in table \ref{table:frolich} for tri-halide perovskites. As a consequence, the formed polarons in these systems will interact with light to form the new states known as exciton-polaron, which are treated in the works of Pollmann, Kane and others \cite{kane,pollmann}. In their works, the exciton Hamiltonian is modified to include the e-ph interaction \blue{by using an energy minimization procedure. This determines an upper bound for the binding energy that depends} on the effective dielectric screening and the LO phonon energy \cite{kane}. 

\begin{figure}[ht] \centering
\includegraphics[scale=0.5]{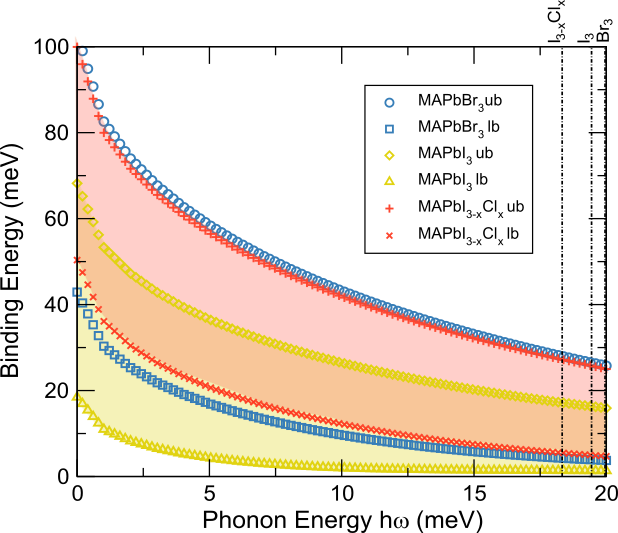}% Here is how to import EPS art
\caption{Exciton binding energy calculated with the Kane model \cite{kane}, for different phonon energies and starting parameters, extracted from \cite{soufiani,sendner,dubey}. These are: $\textrm{MAPbBr}_{3}$-ub ($\varepsilon_{\infty}=4.4$, $\varepsilon_{s}=25.5$,$m_{e}=0.29$, $m_{h}=0.31$), and $\textrm{MAPbBr}_{3}$-lb ($\varepsilon_{\infty}=4.8$, $\varepsilon_{s}=25.5$,$m_{e}=m_{h}=0.15$) in blue symbols; $\textrm{MAPbI}_{3}$-ub ($\varepsilon_{\infty}=5$, $\varepsilon_{s}=19.6$,$m_{e}=0.23$, $m_{h}=0.29$) and $\textrm{MAPbI}_{3}$-lb ($\varepsilon_{\infty}=6.5$, $\varepsilon_{s}=28.8$, $m_{e}=m_{h}=0.12$) in yellow marks;  $\textrm{MAPbI}_{x}\textrm{Cl}_{3-x}$-ub ($\varepsilon_{\infty}=4.5$, $\varepsilon_{s}=24.7$, $m_{e}=0.32$, $m_{h}=0.29$) and $\textrm{MAPbI}_{x}\textrm{Cl}_{3-x}$-lb ($\varepsilon_{\infty}=5.25$, 
$\varepsilon_{s}=29.3$, $m_{e}=0.165$, $m_{h}=0.29$) in red tokens. The abbreviation ub means upper bound, while lb is lower bound. Vertical dashed black lines denotes the LO phonon energy calculated from the linewidth fit.}
\label{fig:polarons} 
\end{figure}

By using the Kane's model, we have reproduced the exciton-polaron binding energies found by Soufiani et al. \cite{soufiani} for $\textrm{MAPbBr}_{3}$ and $\textrm{MAPbI}_{3}$. The procedure consists of settle the parameters of $m_{e}$, $m_{h}$ $\epsilon_{\infty}$ and $\epsilon_{0}$ for each perovskite to then calculate the binding energy with the variable LO phonon energy in the range of  $0$ up to $20$ meV. This is illustrated in figure \ref{fig:polarons} in blue an yellow colors for $\textrm{MAPbBr}_{3}$ and $\textrm{MAPbI}_{3}$, respectively. Likewise, by using an average value computed with the information of $\textrm{MAPbI}_{3}$ \cite{soufiani} and $\textrm{MAPbCl}_{3}$ \cite{sendner}, for the mixed halide perovskite, $\textrm{MAPbI}_{x}\textrm{Cl}_{3-x}$, we arrive to the red colored region show on figure \ref{fig:polarons}. It is worth to mention that the variability of the parameters found in literature has set upper and lower bounds for our binding energy calculations. We can see that the binding energy of the exciton-polaron can acquire different values depending on the phonon branch they are coupling. This, perhaps, could be an explanation of the large difference in binding energies calculated with different experiments as suggested by Soufiani et al. \cite{soufiani}.

Lastly, the LO phonon energies of the tri-halide perovskites obtained in the linewidth section appear in figure \ref{fig:polarons} as the vertical dashed lines. There, the intersections mark the binding energies of the Kane's model with values of $3.86$-$25.91$ meV, $1.63$-$16.52$ meV and $5.65$-$27.39$ meV for $\textrm{MAPbBr}_{3}$, $\textrm{MAPbI}_{3}$ and $\textrm{MAPbI}_{x}\textrm{Cl}_{3-x}$, respectively. Remarkably, these values from the intersections are in agreement with the binding energies retrieved by the EBF model. \blue{Thus, confirming the reliability of our model.} 

\blue{The ability of the EBF model to predict the low values of the exciton binding energy may imply that concept of lattice deformations behind the band-fluctuations potential of eq. (\ref{w}) is mapped into the polaronic contribution for polar and ionic materials.}

\begin{table*}
\caption{\label{table:binding_energy} Binding energies of the first excitonic state of different experimental reports, \blue{retrieved from} the data collected from Jiang et al. \cite{ying}, for GaAs and the tri-halide perovskites.}
\begin{indented}
\item[]\begin{center}
\begin{tabular}{@{}ccccc}
\br
GaAs & Binding Energy & Temperature & Method & Ref.\\ \mr
& 4.2 & 1.2-2.1K & PL/Reflection & \cite{nam}\\
& 4.2 & 1.2K & Absorption & \cite{fehrenbach}\\
& 3.4, 3.2, 3.3, 3, 2.5 & 21K, 55K, 90K, 185K, 294K & Absorption & \cite{sturge}\\
& 4.2 & 1.2 & Absorption (F-sum rule) & \cite{sestu2}\\ \mr
$\textrm{MAPbBr}_{3}$& Binding Energy & Temperature & Method & Ref.\\\mr
 & 40.3 & - & Absorption & \cite{yang}\\
 & 60 & 170-300/80-140K & Absorption (F-sum rule) & \cite{sestu}\\
 & 60 & - & Absorption/PL & \cite{wu}\\
 & 53 & - & Absorption/PL & \cite{mariano}\\
 & 22-25 &100-300K & PL at diff. T & \cite{niesner} \\
 & 36-41 & 5-280K & Absorption (F-sum rule) & \cite{fabian}\\
  & 35-42 & 5-280K & Absorption & \cite{fabian}\\
  & 15.33 & 4.5-150K & PL at diff. T & \cite{tilchin}\\
  & 76 & 4.2K & Magnetoabsorption & \cite{tanaka}\\
  & 25 & 2K & High Field Magnetoabsorption & \cite{galkowski}\\
  &20-40 & 5K & Photoluminescence & \cite{kunugita}\\
  &34-41 & 9-298K & Absorption & \cite{soufiani}\\
  & 64 & - & Absorption & \cite{saba2016}\\
  & 71 & - & Compuational Methods & \cite{bokdam}\\
 \mr
$\textrm{MAPbI}_{3}$& Binding Energy & Temperature & Method & Ref.\\\mr
 & 15 & 50-162K & Absorption &  \cite{even}\\ 
 & 5 & 215-300K & Absorption & \cite{even}\\
 & 37 & 4.2K & Magnetoabsorption & \cite{hirasawa}\\
 & 16 & 8.5-145K & High Field Magnetoabsorption & \cite{miyata}\\
 & 5 & 300K & High Field Magnetoabsorption & \cite{miyata}\\
 & 13 & - & Absorption & \cite{yang}\\
 & 50 & 4.2K & Magnetoabsorption & \cite{tanaka}\\
 & 45 & 78-296K & Absorption & \cite{ishihara}\\
 & 24-32 & 5-280K & Absorption (F-sum rule) & \cite{fabian}\\
 & 22-29 & 5-280K & Absorption & \cite{fabian}\\
 & 34 & 89-140K & Absorption (F-sum rule) & \cite{sestu}\\
 & 29 & 170-300K & Absorption (F-sum rule) & \cite{sestu}\\
 & 25 & 170K and 300K & Absorption & \cite{saba}\\
 & 25-12 & 7-298K & Absorption & \cite{soufiani}\\
 & 9 & 300K & Absorption & \cite{yangostrowski}\\
 & 16 & 2K & High Field Magnetoabsorption & \cite{galkowski}\\
 & 12 & 155-190K & High Field Magnetoabsorption & \cite{galkowski}\\
 & 15-6 & 160-300K & Absorption & \cite{yamada2015}\\
 & 19 & 10-300K & PL at diff. T & \cite{sun}\\
 & 30 & - & Absorption & \cite{koutselas}\\
 & 2 & 4.2K & Revisited Magnetoabsorption & \cite{lin}\\
 & 12.4 & 10-120K & Photocurrent at diff. T & \cite{phuong} \\
 & 45 & - & Computational Methods & \cite{bokdam} \\\mr
$\textrm{MAPbI}_{x}\textrm{Cl}_{3-x}$ & Binding Energy & Temperature & Method & Ref.\\\mr
& 98 & 5-300K & PL at diff. T & \cite{zhang}\\
& 14 & 2K & High Field Magnetoabsorption & \cite{galkowski} \\
& 10 & 190-200K & High Field Magnetoabsorption & \cite{galkowski} \\
& 55($\pm$20) & 160-300K & Absorption & \cite{dinn} \\
& 62 & 160-300K & PL at diff. T & \cite{wukw}\\
  \br
\end{tabular}
\end{center}
\end{indented}
\end{table*}

\section{Conclusions}
We have constructed a new analytical model based on the Elliott equation. In the Elliott model, the coulomb interaction is included in electronic transitions, and, as a consequence, excitons are formed. In this way, the absorption is increased by a set of peaks below the bandgap and a more prominent zone above it due to the Sommerfeld enhanced factor. In the frame of this model, a convolution process is used for the contributions regarding the uncertainty of the measurement and thermodynamic effects. The Lorentzian profile is the preferred for the former while the Gaussian and hyperbolic secant profiles are used for the latter. \blue{Nonetheless, this profiles produce non-analytic functions.}

Motivated by the success of the band-fluctuations model developed in Guerra et al. \cite{guerra}, and extended in Lizarraga et al. \cite{lizarraga}, for describing the fundamental absorption and Urbach tails in direct, indirect and amorphous semiconductors. We have carried the convolution process by using the potential that accounts the fluctuations of the bands. Fluctuations  arise from several factors such as perturbations of thermodynamic origin, impurities or dislocations of the perfect lattice. This produces an analytical function which is coupled with the Lorentzian contribution in a pseudo Voigt profile, similar as the one developed in Soufiani et al. \cite{soufiani}. 

The new Elliott Band-Fluctuations (EBF) model serve us to analyze several  semiconductors such as GaAs and the family of the tri-halide perovskites, $\textrm{MAPbBr}_{3}$, $\textrm{MAPbI}_{3}$, and $\textrm{MAPbI}_{x}\textrm{Cl}_{3-x}$. The fits produce a contribution of $99\%$ \blue{arising from the band-fluctuations term.} The obtained values of exciton binding energy values, bandgap and FWHM are in good agreement with other studies on these materials. 

We then proceeded to analyze the bandgap evolution with temperature to understand the contributions of the thermal expansion and the e-ph interaction. While in some cases the former is negative, as in the case of GaAs, the tri-halide perovskites show a positive character \blue{of it}. For tri-halide perovskites, the dominant contribution for the bandgap arises from thermal expansion expressed as a linear behavior. On the other hand, the negative contribution of the e-ph interaction, which is proportional to the Bose-Einstein distribution, predominates in the bandgap evolution for GaAs. The independent fits, adapted from the models of Varshni and Cardona, accurately predict the Debye energy and thermal expansion coefficient for GaAs and tri-halide perovskites, respectively.

The evolution linewidth retrieved by our model is fitted by using the broadening evolution proposed by Toyozawa and Segall. Here the electron-phonon interaction is divided in the contributions arising from acustic and optical phonons. The former produces a linear term, while the latter build a factor proportional to the bosonic occupation number. And, based on previous studies, we just consider the LO phonon contribution for our materials. These ends up in the determination of the LO phonon energy for the tri-halide perovskites. 

Lastly, we use the LO phonon energies retrieved by our bandgap and linewidth analysis to calculate the exciton binding energy in the presence of the e-ph interaction by using the theory developed by Kane et al. \cite{kane}. Here, we obtain results that lie into the range of the binding energies calculated with our EBF model for the tri-halide perovskites. \blue{Thus, we believe, that given such agreement and correspondence, the EBF model would be useful for experimentalists and for theoretical studies.}

\section*{Acknowledgments}
We would gratefully like to acknowledge the Peruvian National Council for Science, Technology and Technological Innovation (CONCYTEC) for a Ph.D. scholarship under grant no. 236-2015-FONDECYT, the German Academic Exchange Service (DAAD) in conjunction with FONDECYT (grants 57508544 and 423-2019-FONDECYT, respectively), and the Office of Naval Research, Grant No. N62909-21-1-2034

\appendix
\setcounter{section}{1}
\section*{Appendix}
The procedure to arrive to the expression (\ref{eq.trans.cont}) from eq. (\ref{convolcint}) is presented here. First, we write the full expression of eq. (\ref{convolcint}):
\begin{eqnarray}
\langle R^{c}_{cv} \rangle= 2\sqrt{R^{*}}\mathcal{R}_{c}|P_{cv}|^{2} &\int_{E_{g}}^{+\infty} \frac{1}{1-e^{-2\pi\sqrt{R^{*}/(\hbar\omega-E_{g})}}}\times \nonumber \\
&\frac{1}{\sigma}\frac{ e^{(E-\hbar\omega)/\sigma}}{ (1+e^{(E-\hbar\omega)/\sigma} ) ^{2}} dE. 
\end{eqnarray}
%\langle R_{cv} \rangle=&4(2\mu)^{3/2} \pi\sqrt{R^{*}} |M_{cv}|^{2}* \nonumber \\
We first make the substitutions of $x=E-E_{g}$ and $y=x+E_{g}-\hbar\omega=E-\hbar\omega$. The integral becomes:
\begin{eqnarray}
\langle R^{c}_{cv} \rangle= 2\sqrt{R^{*}}\mathcal{R}_{c}|P_{cv}|^{2} \int_{0}^{+\infty} &\frac{1}{1-e^{-2\pi\sqrt{R^{*}/x}}} \times \nonumber  \\
&\frac{1}{\sigma}\frac{ e^{y/\sigma}}{(1+e^{y/\sigma})^{2}} dx. 
\end{eqnarray}
Now, we make the identity $b=2\pi\sqrt{R^{*}}$, and apply integration by parts:
\begin{eqnarray}
u&=\frac{1}{1-e^{-b/\sqrt{x}}} \ \rightarrow \ du=\frac{b x^{-3/2}e^{-b/\sqrt{x}}}{2(1-e^{-b/\sqrt{x}})^2}dx, \nonumber\\
dv&=\frac{1}{\sigma}\frac{ e^{y/\sigma}}{(1+e^{y/\sigma})^{2}} \ \rightarrow \ v=-\frac{1}{(1+e^{y/\sigma})}.
\end{eqnarray}
The integral in eq. (\ref{convolcint}) becomes: 
\begin{eqnarray}
I = \frac{1}{1-e^{-b/\sqrt{x}}} &\frac{-1}{1+e^{y/\sigma}} \Bigg|_{0}^{+\infty} - \nonumber\\
&\int_{0}^{+\infty} \frac{-1}{1+e^{y/\sigma}} \frac{b x^{-3/2}e^{-b/\sqrt{x}} }{2(1-e^{-b/\sqrt{x}})^2} dx,  \end{eqnarray}
\begin{eqnarray}
 I=&\frac{1}{1+e^{(E_{g}-\hbar\omega)/\sigma}} + 
 \frac{b}{2} \int_{0}^{+\infty} dx \frac{1}{\sqrt{x^{3}}} \times \nonumber \\
 &\left(  \frac{1}{e^{b/2\sqrt{x}}-e^{-b/2\sqrt{x}}} \right)^{2} \left( \frac{1}{1+e^{(x+E_{g}-\hbar\omega)/\sigma}}   \right). \label{integralhuge}
\end{eqnarray}
This can be solved by using the Sommerfeld expansion, i.e,
\begin{eqnarray}
\int_{0}^{+\infty}& H(x) \frac{1}{1+e^{(x-\mu)/\sigma}} dx = \nonumber \\
&\int_{0}^{\mu} H(x)dx  + \frac{\pi^2}{6} \sigma^2 H'(x) \bigg|_{x=\mu} + O\left(\frac{\sigma}{\mu}\right)^4. \label{somexp}
\end{eqnarray}

Here, the variable $\mu$ is $\hbar\omega - E_{g}$.  Likewise, the function $H(E)$ (see eq. \ref{he}) satisfies the two conditions for the expansion, it does not grow faster than polinomially in $x$ when $x$ goes to infinity, and, it vanishes in the limit of $x$ approaching to zero.
\begin{equation}
H(x)=\frac{1}{\sqrt{x^{3}}}\left( \frac{1}{e^{b/2\sqrt{x}}-e^{-b/2\sqrt{x}}} \right)^{2}. \label{he}
\end{equation}

We neglect higher order terms, $\sigma^2$ ,in the Sommerfeld expansion, i.e,
\begin{eqnarray}
    \int_{0}^{+\infty}H(x)& \frac{1}{1+e^{(x-\mu)/\sigma}} dx = \nonumber \\
    &\int_{0}^{\mu} H(x) dx = \frac{2}{b\left( e^{b/\sqrt{\hbar\omega-E_{g}}} -1 \right)}.\label{firstterm}
\end{eqnarray}

Lastly, we replace eq. (\ref{firstterm}) on eq. (\ref{integralhuge}), and the transition rate for the continuum is:
\begin{eqnarray}
\langle R^{c}_{cv} \rangle=2\sqrt{R^{*}}\mathcal{R}_{c}|P_{cv}|^{2}& \Bigg( \frac{1}{1+e^{(E_{g}-\hbar\omega)/\sigma}}+ \nonumber\\
&\frac{1}{e^{2\pi\sqrt{R^{*}/(\hbar\omega-E_{g}})}-1} \Bigg).
\end{eqnarray}

\section*{References}
\bibliography{iopart-num}

\end{document}